\documentclass{article}
\pdfoutput=1

\usepackage{geometry}

\usepackage{cmap}						
\usepackage[T2A]{fontenc}				
\usepackage[utf8]{inputenc}				
\usepackage{csquotes}
\usepackage[english,russian]{babel}	

\usepackage{indentfirst} 

\usepackage{titlesec}

\usepackage{amsmath}
\usepackage{amsfonts}
\usepackage{amssymb}
\usepackage{stmaryrd} 
\usepackage{bbold} 

\IfFileExists{subcaption.sty}{
	\usepackage{subcaption}
	\captionsetup{compatibility=false}
}{
	\usepackage{caption}
	\DeclareCaptionSubType{figure}
	\makeatletter
	\caption@For{subtypelist}{%
		\newenvironment{sub#1}%
		{\caption@withoptargs\subcaption@minipage}%
		{\endminipage}}%
	\newcommand*\subcaption@minipage[2]{%
		\minipage#1{#2}%
		\setcaptionsubtype\relax}
	\makeatother
}

\usepackage{xspace}

\usepackage{tikz}
\usetikzlibrary{calc,positioning}

\usepackage{textgreek}

\usepackage[
	style=gost-numeric,
	language=auto,
	babel=other,
	backend=biber,
	sorting=ntvy,
	url=false,
	doi=false,
	isbn=false,
	movenames=false,
	maxnames=100
]{biblatex}
\DeclareSourcemap{
	\maps[datatype=bibtex,overwrite=false]{
	 \map{
			\step[fieldset=language,fieldvalue=english]
	 }
	\map{
		\step[fieldsource=language,match=\regexp{English},replace=\regexp{english}]
	}
	\map{
		\step[fieldsource=language]
		\step[fieldset=hyphenation,origfieldval]
	}
	}
}
\NewBibliographyString{langjapanese}
\NewBibliographyString{fromjapanese}

\usepackage{multirow}
\usepackage{rotating}
\usepackage{array}
\usepackage{pbox}

\bibliography{/home/leshyk/Science/library.bib}

\usepackage{setspace}
\usepackage{textgreek}
\usepackage{rotating}
\usepackage{pbox}

\newcommand{\SDuTE}{\textbf{ОПоВ}\xspace}

\newcommand{\EE}[1]{\mathbb{E}[#1]}
\newcommand{\PP}[1]{\mathbb{P}[#1]}

\usepackage{lineno}

\newcommand{\chapterEnTitle}{}
\newcommand{\chapterRuTitle}{}
\newcounter{mychapternumber}

\begin{document}
	


\renewcommand{\chapterEnTitle}{
  Machine learning for subgroup discovery under treatment effect
} 

\renewcommand{\chapterRuTitle}{
    Машинное обучение для выявления подгрупп индивидов, сильно реагирующих на воздействие
}          


\setcounter{mychapternumber}{0} 

\hyphenation{Diag-nos-tic-Tests-Sca-ling-And-In-fer-ring длинное-название-возможно-например-на-немецком long-title-possible-for example-in-German} 

\newcommand{\addtocen}[2]{}
\newcommand{\addtocru}[2]{}
\newenvironment{abstr}{
}{
}
\newcommand{\mailtoMLABSEDauthor}[3]{\textit{#1}}
\newcommand{\abstractnameENG}{\bf Abstract}
\newcommand{\keywordsENG}{\bf Keywords}
\newcommand{\acknowledgementsENG}{\bf Acknowledgements}
\renewcommand{\abstractname}{\bf Аннотация}
\newcommand{\keywords}{\bf Ключевые слова}
\newcommand{\acknowledgements}{\bf Благодарности}
\newcommand{\delnewpagebeforech}{\newpage}
\makeatletter
\def\chapter{\@ifstar\@chapter\@@chapter}
\def\@chapter#1{
	\begin{center}
		\Huge \chapterRuTitle
	\end{center}
}
\def\@@chapter#1{
}
\makeatother
\newcommand{\thechapter}{}
\newcommand{\FloatBarrier}{}
\renewcommand\newrefcontext[1][tmp]{}
{%
	\cleardoublepage
	\let\clearpage\relax
	\begin{center}
		\Huge \chapterEnTitle
	\end{center}
	\addtocen{chapter}{\chapterEnTitle} 
	\addtocru{chapter}{\chapterRuTitle} 
}%

\begin{abstr}
  Aleksey Vladimirovich Buzmakov, PhD in Computer Science, Cand. of Sci. in Engineering, Senior Research Fellow, National Research University Higher School of Economics, 37a Gagarina Bulvar, Perm, Russia, %
  \mailtoMLABSEDauthor{avbuzmakov@hse.ru}{Dear Author}{avbuzmakov@hse.ru}.    
  \par
  {\normalfont \abstractnameENG.} 
  In many practical tasks it is needed to estimate an effect of treatment on individual level. For example, in medicine it is essential to determine the patients that would benefit from a certain medicament. In marketing, knowing the persons that are likely to buy a new product would reduce the amount of spam. In this chapter, we review the methods to estimate an individual treatment effect from a randomized trial, i.e., an experiment when a part of individuals receives a new treatment, while the others do not. Finally, it is shown that new efficient methods are needed in this domain.
  \par
  {\normalfont \keywordsENG.} SuDiTE, Subgroup Discovery under Treatment Effect, Treatment Effect Estimating, Randomized Experiment
  \par
  {\normalfont \acknowledgementsENG.} The publication was prepared within the framework of the Academic Fund Program at the National Research University Higher School of Economics (HSE) in 2019 (grant \textnumero~19-04-048) and by the Russian Academic Excellence Project ``5-100''. The author of the chapter is grateful to Xenia A. Naidenova, Vladimir Parkhomenko, Evgeniya Shenkman, and the chapter reviewers for a lot of fruitful comments.
  \delnewpagebeforech 

  \chapter*{\normalsize\thechapter. \chapterRuTitle} 
  
  Алексей Владимирович Бузмаков, PhD, к.т.н., с.н.с., Научно-исследовательский унивеситет Высшая школа экономики, Россия, Пермь, бульвар Гагарина 37а,
  \mailtoMLABSEDauthor{avbuzmakov@hse.ru}{Dear Author}{avbuzmakov@hse.ru}. 
  \par
  {\normalfont \abstractname.} 
  Оценка эффекта от воздействия на индивидуальном уровне необходима во многих областях знаний от медицины до маркетинга. Действительно, общество выиграет, если будет возможность определять, на кого подействует какое-либо лекарство. А при отправлении рассылки только тем людям, которых интересует конкретный товар, уменьшится количество спама и снизятся издержки рекламной кампании. В главе рассматриваются существующие методы оценки эффекта от воздействия на индивидуальном уровне и показывается на основе компьютерного эксперимента необходимость создания новых эффективных методов в этой области знаний. 
  \par
  {\normalfont \keywords.} Оценка эффекта от воздействия, Выделение подгрупп, Случайный эксперимент, Ансамблевые методы, Машинное обучение
  \par
  {\normalfont \acknowledgements.} Публикация подготовлена в ходе проведения исследования (\textnumero~19-04-048) в рамках Программы <<Научный фонд Национального исследовательского университета <<Высшая школа экономики>> (НИУ ВШЭ)>> в 2019 г. и в рамках государственной поддержки ведущих университетов Российской Федерации ``5--100''. Автор главы выражает благодарность К.А. Найденовой, В.А Пархоменко, Е.А. Шенкман и рецензентам главы за большой количество ценных замечаний и комментариев.

\end{abstr}

\begin{refsection}[my_folder/my_biblio.bib] 
  
\newrefcontext[labelprefix=\thechapter.] 



\section*{Введение} 
\addtocru{section}{Введение} 
\addtocen{section}{Introduction} 

Мы влияем на окружающий нас мир, а мир, в свою очередь, влияет на нас. Каждое воздействие с нашей стороны преследует некоторую цель. Но как мы можем оценить эффективность этого воздействия? Этот вопрос становится особенно актуальным, когда наше воздействие комбинируется с другими, внешними воздействиями. Например, если мы хотим проверить новое лекарство для понижения артериального давления, нам необходимо отделить эффект лекарства от внешних причин таких, как изменение погоды или изменение настроения пациента. При проверке эффективности новой маркетинговой стратегии нужно отделить эффект самой стратегии от конъюнктурных изменений, a при тестировании нового алгоритма поиска информации --- эффект самого алгоритма от изменений в поведении потребителей поискового продукта. 

Для решения задачи оценки эффективности того или иного воздействия применяют <<слепое>> испытание, называемое также A/B-тестированием~\cite{Taddy2016}. Идея слепого испытания заключается в следующем. Пусть нам нужно протестировать новое лекарство. Тогда пациенты, согласившиеся участвовать в испытании, будут поделены на две группы: тестовую и контрольную (или плацебо-группу). При этом сами пациенты, а также врачи, обследующие пациентов, не должны знать, к какой группе относятся пациенты. Во время эксперимента, пациентам из тестовой группы будут давать новое лекарство, а пациентам из контрольной группы --- плацебо, вещество без лекарственных свойств. В дальнейшем, при сравнении результатов измерений контрольных параметров, например, артериального давления, у пациентов в контрольной и тестовой группах можно выявить, оказало ли тестируемое лекарство воздействие, отличающееся от эффекта плацебо. Для количественной оценки эффективности воздействия применяют различные статистические тесты. В частности, распространённым подходом является применение теста Стьюдента для проверки наличия значимой разницы в средних у наблюдаемого параметра для тестовой и контрольной групп.

Такой подход позволяет заключить с некоторой степенью уверенности об эффективности того или иного воздействия: эффективно ли новое лекарство; успешна ли маркетинговая компания. Однако из опыта мы знаем, что лекарства по разному влияют на разных людей. Тоже самое мы можем наблюдать и в маркетинге. Часть людей будут рады получить скидку, а другая часть --- раздражена оказанием давления на них. Ещё более ярким этот эффект наблюдается при удержании клиентов. Для некоторых клиентов, которые находятся в зоне риска, может существенно увеличиться вероятность их ухода к конкурентам после воздействия, которое должно было уменьшить эту вероятность~\cite{Radclifte2008,Ascarza2018}.

Более того, воздействие, которое при обычном подходе рассматривается как неэффективное, является не эффективным в среднем. Однако не исключено, что есть подгруппы индивидов, на которых исследуемое воздействие оказало положительный результат. И тогда, ресурсы, использованные на разработку  и проведение такого воздействия, потрачены не напрасно. Относительно молодая область интеллектуальной обработки данных, называемая выявлением однородных подгрупп объектов по отношению к воздействию\footnote{
  \textit{англ.} subgroup discovery under treatment effect
} (\SDuTE), посвящена именно таким вопросам. Под объектами здесь очень часто понимаются индивиды, однако возможны практические приложения, в которых выделяются группы других сущностей, например, таких как города~\cite{Asher2016}.


В данной главе мы рассмотрим основные подходы к выделению \SDuTE методами машинного обучения. Такие методы в большинстве своём строят деревья решений, которые состоят из корня дерева, внутренних вершин и листьев (терминальных вершин). Каждая вершина соответствует множеству исследуемых объектов. С каждой вершиной, за исключением листовых, связан вопрос к соответствующим объектам. Такой вопрос делит объекты на подгруппы и эти подгруппы соответствуют вершинам-детям для рассматриваемой вершины. Например, имея выборку пациентов, мы можем разбить эту выборку на мужчин и женщин, которые станут вершинами-детьми для корневой вершины. Таким образом, при правильном подборе вопросов в вершинах, листья дерева могут соответствовать подгруппам пациентов, которые получают максимальный эффект при приёме лекарства. В этой главе мы рассмотрим различные подходы к подбору вопросов в вершинах, чтобы образованные подгруппы объектов были однородны с точки зрения эффекта от  воздействия.

Нужно отметить, что подходы к выделению \SDuTE не ограничиваются построением одного или нескольких деревьев решений. В эконометрике было разработано большое количество подходов к выделению \SDuTE на основании структурных моделей, определённого рода выражений с неизвестными коэффициентами, основанных на знаниях предметной области. В этих подходах большое вниманию уделено проблеме <<само-отбора>>, то есть когда объекты разбиваются на контрольную и тестовую выборку неслучайным образом. Такая проблема часто возникает в социальных науках, когда у исследователя нет возможности назначать воздействия и люди сами определют подвергаться воздействию или нет. Например, очень большое количество таких методов посвящено проблеме оценки эффективности программ, помогающих в поиске работы в США~\cite{Imbens2009,Card2010}. Люди, которые проходят такие программы, вероятно существенно отличаются от тех, кто их не проходит. 
Интересующийся читатель может проконсультироваться со специализированными обзорами~\cite{Card2010,Linden2016}. Целью данной главы является анализ методов машинного обучения, приложимых к проблеме выделения \SDuTE, предназначенных в большинстве случаев для исследования результатов случайного эксперимента.

Прежде чем продолжить, нужно остановиться на одном важном моменте. Ряд исследователей подчёркивает, что методы выделения \SDuTE могут  порождать ложные результаты~\cite{Assmann2000,Pocock2002,Brookes2004,Rui2007}. Действительно, подобные методы в явном или неявном виде предполагают множественные сравнения различных статистик для контрольной и тестовой групп. Таким образом, выделение той или иной подгруппы индивидов конкретным методом, может быть отнесено на случайное стечение обстоятельств. Поэтому, подгруппы, найденные каким-либо методом, должны рассматриваться как гипотезы, каждая из которых требует дальнейшей проверки при проведении дополнительного эксперимента.

Глава организована следующим образом. В первом разделе формулируются задачи оценки эффекта от воздействия (далее --- задача 1) и выявления подгрупп, однородных с точки зрения эффекта от воздействия (далее --- задача 2). Во втором разделе обсуждается вопрос оценки качества обученных моделей для оценки эффекта от воздействия. В силу специфики рассматриваемой задачи эта оценка не является тривиальной. В третьем разделе рассматриваются некоторые из существующих методов, сгруппированных в 4 блока в зависимости от типа построения модели оценки эффекта от воздействия. В этом же разделе приведены результаты численного эксперимента, показывающие, что даже лучшие подходы ещё далеки от безупречной работы и что требуются новые более эффективные методы оценки эффекта от воздействия.

\section{Задачи выделения \SDuTE и оценки эффекта от воздействия}
\addtocru{section}{Задачи выделения \SDuTE и оценки эффекта от воздействия} 
\addtocen{section}{Individual Treatment Effect Estimation Task} 
\subsection{Обозначения}
\addtocru{subsection}{Обозначения} 
\addtocen{subsection}{Preliminaries} 
Для формулировки рассматриваемой в данной главе задачи необходимо задать: объекты (наблюдения) и их описания, набор воздействий, оказываемых на объекты, а также каким образом измеряется реакция объектов на эти воздействия. Рассмотрим эти составляющие задачи подробнее.

Пусть есть множество объектов $G$, каждому из которых соответствует некоторое описание из множества $D$, заданное посредством функции $\delta:G\rightarrow D$. Например, множество объектов $G$ может состоять из пациентов, в то время как описание каждого пациента (элемент множества $D$) может включать его пол, возраст, вес, данные анализов, историю болезни и другое. Таким образом множество $D$ будет состоять из кортежей со значениями этих характеристик. Если в задаче рассматриваются только $K$ численных характеристик, то будем эти характеристики обозначать через $X_i$, $i \in \overline{1,K}$. В этом случае элементы $D$ примут вид $<x_1,x_2,\dots,x_K>$, где $x_i$ -- это значение характеристики $X_i$.
Для описания объекта $g \in G$ введём обозначение $d_g=\delta(g)$.

Предположим, что на объект могло быть оказано одно воздействие из множества $\mathtt{T}$. Воздействие $\bot \in \mathtt{T}$ соответствует некоторому базовому воздействию, с которым требуется сравнить все остальные воздействия. Часто таким базовым воздействием является отсутствие воздействия. Например, при тестировании лекарственного препарата множество $\mathtt{T}$ обычно состоит из двух воздействий: каждому пациенту дают либо лекарственное средство, либо плацебо. При этом базовым является именно плацебо, так как в ходе такого тестирования исследователь хочет знать является ли лекарственное средство эффективным по сравнением с плацебо. Альтернативой может служить сравнение нового лекарственного препарата и некоторого старого. В этом случае лечение старым лекарственным препаратом является базовым воздействием. С другой стороны в маркетинге при разработке маркетинговой кампании базовым воздействием является отсутствие какого-либо воздействия, в то время как другие воздействия из $\mathtt{T}$, будут предоставлять, например, скидку клиенту.

Оказанное воздействие для каждого объекта задаётся посредством функции $t: G \rightarrow \mathtt{T}$. Также для упрощения последующей записи обозначим $t_g=t(g)$. Также отметим, что $t_g$ можно рассматривать как случайную величину, которую будем обозначать как $T$, наряду с $Y$ и $X_i$, в тех ситуацию, когда будет рассматриваться сам процесс получения наблюдений вида $\left<y_g,t_g,x_1^g,x_2^g\right>$.

В рамках рассматриваемой задачи полагаем, что оказанное воздействие может влиять на некоторый целевой параметр $Y$. Например, таким параметром при тестировании лекарственного препарата для лечения диабета II типа может быть средний уровень сахара у пациента в последующую неделю после эксперимента, а в области маркетинга -- общий объём выручки с одного клиента за месяц после начала кампании.
Здесь стоит отметить, что значение этого параметра зависит от конкретного объекта $g$, оказанного воздействия $\mathtt{t} \in \mathtt{T}$ и, в общем случае, от внешней среды $E$ в момент оказания воздействия. Таким образом, целевой параметр $Y$ мы будем считать функцией $Y(g,\mathtt{t},E)$, где $g \in G$ и $\mathtt{t} \in \mathtt{T}$. Последнюю составляющую ($E$) мы не будем явно рассматривать в дальнейшем, так как при случайном распределении воздействий, многие случайные вариации внешней среды лишь увеличивают уровень шума в данных. Однако следует помнить, что если эксперимент проводился в среде $E$, отличной от реальной, результаты такого эксперимента должны осторожно переноситься на реальный случай. Подробное обсуждение этой проблемы может быть найдено в~\cite{Deaton2010,Glennerster2017,Athey2017}.
Далее будем считать, что $Y$ это функция $Y(g,\mathtt{t})$, которую для удобства мы также можем обозначать как $Y_g(\mathtt{t})$. 

Стоит отметить, что после проведения эксперимента для некоторых объектов из $G$ известно значение целевого параметра, но только для одного воздействия: $Y^{(exp)}_g=Y(g,t_g)$, где $t_g$ -- то воздействие, которому мы подвергли объект $g \in G$.
Ни для какого объекта $g$ невозможно наблюдать значение целевой переменной $Y$ при нескольких воздействиях одновременно. В литературе данное ограничение называется фундаментальной проблемой выявления причинности~\cite{Holland1986}\footnote{\textit{англ.} the fundamental problem of casual inference}. Именно эта фундаментальная проблема является одной из основных проблем при работе с задачей выявления \SDuTE. 

%
\subsection{Формальная постановка задачи}
\addtocru{subsection}{Решаемая задача} 
\addtocen{subsection}{Task Description} 
Пусть $G_0$ --- генеральная совокупность объектов. В данном случае под генеральной совокупностью мы будем понимать потенциальное множество всех объектов, по отношению к которым мы хотели бы принимать решение о том или ином воздействии на основании в том числе рассматриваемого случайного эксперимента. Например, при разработке нового лекарственного препарата генеральной совокупностью, будут все люди, как ныне живущие, так и будущие. В случае же маркетинговой кампании, под генеральной совокупностью, можно понимать всех текущих и будущих клиентов. Таким образом, генеральная совокупность $G_0$ как правило ненаблюдаема целиком. Мы можем наблюдать лишь её некоторое подмножество $G \subseteq G_0$. Однако выводы анализа случайного слепого исследования мы хотим получить на всю генеральную совокупность $G_0$.

Целью анализа случайного слепого исследования является вывод о величине эффекта $\Delta Y_g(\mathtt{t}) = \tau(g,\mathtt{t})$ от воздействия $\mathtt{t} \in \mathtt{T}$ для каждого объекта генеральной совокупности $g \in G_0$ (для удобства обозначим $\tau_g(\mathtt{t})=\tau(g,\mathtt{t})$). Здесь под величиной эффекта $\tau(g,\mathtt{t})$ понимается разница в целевом параметре под воздействием и без него, т.е. $\tau(g,\mathtt{t})=Y_g(\mathtt{t}) - Y_g(\bot)$. Так, например, чтобы посчитать точный эффект от некоторого лекарственного препарата $\mathtt{t}$ для пациента $g$, нам бы пришлось \emph{одновременно} и дать пациенту препарат $\mathtt{t}$, чтобы померить $Y_g(\mathtt{t})$ и дать плацебо, чтобы померить $Y_g(\bot)$, что невозможно.

Получаем, что точно посчитать эффект от воздействия не представляется возможным. Это принципиально отличает задачу выявления \SDuTE от классических задач машинного обучения: классификации и регрессии. Таким образом, нашей целью будет оценка эффекта от воздействия $\hat{\tau}(g,\mathtt{t})$ для объекта $g$ и воздействия $\mathtt{t}$. Такую оценку можно построить только по нескольким объектам, ``похожим'' на объект $g$. Именно для определения этой похожести было упомянуто, что каждому объекту $g$ соответствует его описание $\delta(g) \in D$, которое, например, для пациентов может включать его историю болезни, а для клиентов описывать его средний чек и частоту покупок.

С этой точки зрения мы будем считать неразличимыми объекты, имеющих одно описание, т.е. такие $g_1$ и $g_2$, что $\delta(g_1) = \delta(g_2)$. Однако для этих объектов эффекты от воздействия могут быть разными, т.е. вероятно, что $\tau(g_1,\mathtt{t}) \neq \tau(g_2,\mathtt{t})$. Действительно, если, например, мы не контролируем на диету пациента (описание пациента не содержит информации о его диете), то размер эффекта лекарственного может существенно отличаться в зависимости от диеты пациентов. Таким образом, вместо анализа точного значения эффекта для каждого объекта $g$, нам необходимо перейти к анализу распределения эффектов от воздействия для похожих объектов, $\tau_g(\mathtt{t}) \sim f_{\tau(\mathtt{t})}(d=\delta(g))$. Так, например, если описание пациента содержит только информацию о его возрасте, то вместо расчёта точного значения эффекта от воздействия $\mathtt{t}$ для пациента $g$, нам будет необходимо исследовать распределение эффектов от воздействия для пациентов такого же возраста $\delta(g)$, как пациент $g$.

На практике анализ распределений эффектов от воздействия $\tau_g(\mathtt{t}) \sim f_{\tau(\mathtt{t})}(d=\delta(g))$ сводится к одной из двух задач. Первая задача состоит в оценке условного среднего эффекта от воздействия для этого распределения, что часто необходимо для более тонкой настройки воздействия. Например, если от маркетинговой кампании для клиента $g$ ожидается большой эффект, то можно дать клиенту большую скидку, чем если ожидаемый эффект маленький. Формально такая оценка эффекта от воздействия может быть записана следующим образом:
\begin{equation}\label{eq:task1}
  \hat{\tau_g}(\mathtt{t})=\EE{f_{\tau(\mathtt{t})}(d=\delta(g))} = \EE{Y_g(\mathtt{t}\mid \delta(g)=d)-Y_g(\bot\mid \delta(g)=d)}.
\end{equation}

Если рассматриваемые объекты не различимы, то есть для любых двух объектов $g_1$ и $g_2$ их описания одинаковые $\delta(g_1)=\delta(g_2)$ (мы ничего не знаем про наши наблюдения кроме переменной отклика и оказанного воздействия), то задача оценки условного среднего эффекта от воздействия сводится к простой задаче нахождения среднего эффекта от воздействия на всей генеральной совокупности $G_0$.

Однако для многих практических задач может потребоваться определить области пространства описаний $D_{\tau} \subseteq D$, в которых распределение эффекта от воздействия имеют определённые свойства. В частности, важным свойством условного распределения является статистически значимое отличие от нуля среднего (медианного) значения этого распределения.  Например, пусть нас интересует группа пациентов, на которую тестируемое лекарство оказывает положительный эффект. Тогда требуется найти подмножество пространства признаков $D_{\tau}$ такое, что 
\begin{equation}\label{eq:task2}
(\forall d \in D_{\tau})\mathbb{P}(\tau(\mathtt{t}\mid d) > 0) > \alpha,
\end{equation}
где $\alpha$ -- некоторый уровень значимости, обычно берущийся как $95\%$ или $99\%$. 

Вторая задача является задачей выявления \SDuTE, т.е. подгрупп, однородных с точки зрения эффекта от воздействия, где $\{g \in G \mid \delta(g) \in D_{\tau}\}$ является подгруппой, а требуемые свойства (например, статистически значимое отличие среднего значения распределений от нуля) соответствующих условных распределений обуславливают однородность этой подгруппы. На практике такую подгруппу удобно представлять как множество подгрупп, объединение которых совпадает с $\{g \in G \mid \delta(g) \in D_{\tau}\}$, каждую из которых можно проинтерпретировать. Действительно, если мы обнаруживаем, что по каким-то причинам, тестируемое лекарство хорошо работает либо на пациентах в возрасте от 25 до 40 лет, либо на женщинах старше 16 лет, то с точки зрения интерпретации проще разбить эту подгруппу пациентов на две: женщины старше 16 и мужчины в возрасте от 25 до 40.

Задачи оценки эффекта от воздействия (первая задач) и задача выявления \SDuTE неразрывно связаны между собой. Поэтому в этой главе мы рассматриваем обе задачи и методы, которые умеют их решать. 

%
Отдельно необходимо отметить, что решение задач, сформулированных выше, основывается на результатах случайного слепого тестирования, что подразумевает наличие выборки данных определённого вида. Такая выборка включает в себя: выборку объектов $G \subset G_0$, полученных из генеральной совокупности; функцию распределения объектов по воздействиям $t:G\rightarrow \mathtt{T}$; значение целевой переменной $Y(g,t(g))$ для каждого объекта выборки $g\in G$, а также функцию соответствия объектов $G$ их описаниям ($\delta: G \rightarrow D$). Таким образом, наши данные должны быть устроены следующим образом. Каждое наблюдение в данных (строка таблицы) является объектом $g$, а характеристиками наблюдений, помимо собственного самих описаний $\delta(g)$ объектов, должны быть также тип воздействия $t(g)$, оказанного на объект, и значение целевой переменной для этого объекта под этим воздействием $Y_g(t(g))$. Далее, на основании этих данных нам необходимо решать задачи (\ref{eq:task1}) и (\ref{eq:task2}) для выделения \SDuTE на всей генеральной совокупности.

\section{Проверка качества работы подходов}
\addtocru{section}{Проверка качества работы подходов} 
\addtocen{section}{Model Verification} 
Прежде, чем перейти к обзору самих методов по выделению \SDuTE рассмотрим, каким образом может быть проверено качество работы того или иного подхода.
Здесь нужно отметить, что методы выделения \SDuTE не относятся к методам обучения без учителя, так как некоторая информация по отношению к какой группе следует отнести объект содержится в целевом признаке $Y$. Однако относить методы по выделению \SDuTE к методам обучения с учителем также некорректно. Действительно, истинное значение эффекта от воздействия для индивидуального объекта принципиально ненаблюдаемо, для каждого объекта известно лишь значение целевого признака при каком-то одном воздействии.  Эта особенность методов выделения \SDuTE накладывает определённые ограничения на возможность проверки качества результата того или иного метода. 

\subsection{Проверка качества работы подходов на синтетических данных}
\addtocru{subsection}{Проверка качества работы подходов на синтетических данных данных} 
\addtocen{subsection}{Model Verification on Syntactic Data} 

Действительно, если мы знаем, что объект отнесён, например, к группе с положительным эффектом от воздействия, то как можно это проверить? Самым простым и часто используемым способом является тестирование некоторого метода на синтетических данных. Для этого в явном виде задаётся зависимость переменной отклика $Y$ от воздействия $\mathtt{t}$ и независимых переменных $X_i$ и далее порождаются данные по этому закону. 

  Рассмотрим следующий пример. Пусть есть множество наблюдений $G$.
  Пусть множество описаний этих объектов состоит из кортежей на
  значениях двух числовых признаков $d=<x_1,x_2>$, и существует всего
  одно воздействие ($\mathtt{t}=1$) помимо базового. При этом для
  простоты считаем, что для контрольной выборки воздействие
  соответствует 0. Пусть наблюдения соответствуют случайными
  точками в двухмерном пространстве $<x_1,x_2>$ (для примера пусть они
  взяты из стандартного нормального двухмерного закона распределения).
  Пусть тогда случайное наблюдение $g$ соответствует точке
  $\delta(g)=\delta_g=<x_1^g,x_2^g>$. Также, пусть в рамках
  случайного эксперимента на объект $g$ мы оказали воздействие
  $t_g \in \{0,1\}$, только одно для каждого наблюдения $g$. Тогда для
  каждого наблюдения $g$ мы можем в явном виде задать зависимость
  между переменной отклика $Y_g(t_g)$, и описаниями $\delta(g)=\left<x_1^g,x_2^g\right>$:
  \begin{equation}
    Y_g(t_g,x_1^g,x_2^g)=t_g\cdot x_1^g+x_2^g+\epsilon^g \label{eq:synt-data}
  \end{equation}
  где $\epsilon^g$ имитирует случайный шок, т.е. такую случайную
  величину, которая меняет точную функциональную зависимость между
  $Y_g(t_g)$ и $\delta_g$ на стохастическую, такая случайность нужна
  для того, чтобы смоделировать влияние ненаблюдаемых факторов. Чем больше
  случайный разброс $\epsilon$, тем менее явно в данных наблюдается
  зависимость между $\delta_g=\left<x_1^g,x_2^g\right>$ и $Y_g(t_g)$.
  
  Пусть $X_1$, $X_2$, $\epsilon$ нормально-распределенные случайные величины,
  а назначение воздействия $t_g \in \{0,1\}$ равновероятно для базового воздействия 0, и тестируемого воздействия 1.
  В таблице~\ref{tbl:synt-data} показаны некоторые наблюдения, которые могли бы находиться в токой синтетической выборки данных. В первой колонке находится уникальный идентификатор наблюдения. Далее идут 4 колонки, содержащие информацию о каждом наблюдении. В последней колонке $\epsilon_g$ приводится уровень индивидуального шока для каждого наблюдения. При этом это значение используется только для порождения $Y_g$, но не используется при обучении тестируемой модели, так как это ненаблюдаемый шок. Так при обучении для первого наблюдения мы будем знать, только, что для него оказывалось тестируемое воздействие ($t_g=1$), что переменная отклика $Y_g(t_g=1)=-1.12$, и при этом для первого наблюдения известны две характеристики $x_1^{g_1}=-0.9$ и $x_2^{g_2}=-1.56$. Соответственно колонка $\epsilon^g$ при этом неизвестна и приведена в таблице \ref{tbl:synt-data} только для демонстрации работы зависимости (\ref{eq:synt-data}). Действительно, в соответствии с законом (\ref{eq:synt-data}).
  \begin{equation*}
   -1.12 = 1 \cdot (-0.9) -1.56 + 1.34
  \end{equation*}
  
  \begin{table}
    \caption{Пример наблюдений синтетической выборки данных, в которой два воздействия 0 и 1, две независимых переменных $X_1$ и $X_2$, а также связь переменной отклика $Y$ с этими переменными, заданная условием (\ref{eq:synt-data}).}
    \label{tbl:synt-data}
    \centering
    \begin{tabular}{l||c|c|cc||c}
      \hline
      \hline
      &
      $Y_g$ & $t_g$ & $x_1^{(g)}$ & $x_2^{(g)}$  & $\epsilon_g$ \\
      \hline
      $g_1$ &
      -1.12 & 1 & -0.9 & -1.56 & 1.34 \\
      $g_2$ &
      -0.48 & 1 & 1.02 & -1.07 & -0.43 \\
      $g_3$ &
      0.35 & 1 & 0.66 & -0.14 & -0.17 \\
      $g_4$ &
      0.99 & 0 & -0.64 & 0.1 & 0.89 \\
      $g_5$ &
      -1.01 & 0 & 0.49 & -0.41 & -0.6 \\
      $g_{6}$ &
      -0.34 & 0 & 1.42 & 0.38 & -0.72\\
      \hline
      \hline
    \end{tabular}
  \end{table}

  Зная закон (\ref{eq:synt-data}), мы можем породить данные любого размера и можем также контролировать уровень случайного шока $\epsilon^g$ как по отношение к переменной отклика $Y_g(t_g)$, так и по отношению к размеру эффекта $Y_g(\mathtt{t}=1) - Y_g(\mathtt{t}=0)$. Более того, когда данные порождаются таким образом, требуемые \SDuTE заранее
  известны. Действительно, для данных, порождённых в соответствии с законом
  (\ref{eq:synt-data}) при $X_1 > 0$, будет наблюдаться положительный
  эффект от воздействия, а при $X_1 \leq 0$ --- отрицательный.
  Соответственно можно проверить, куда отнесено каждое наблюдение тестируемым методом и посчитать точность (какая часть наблюдений в найденных \SDuTE, действительно относится к \SDuTE с точки зрения закона порождения данных), полноту (какая часть наблюдений, относящихся к \SDuTE, попала хотя бы в один найденный \SDuTE), корреляции рассчитанного и истинного эффекта от воздействия, а также другие характеристики качества. 
  
  Синтетические данные для тестирования методов используются во многих работах~\cite{Cai2010,Tian2014,Guelman2012,Guelman2015a,Guelman2015,Lipkovich2011,Burke2014,Athey2016}. В качестве второго примера, рассмотрим один из законов порождения синтетических данных, используемых в работе~\cite{Athey2016}:
  \begin{equation}
    \label{eq:synt-data2}
    Y_g(t_g,\delta_g=\left<x_1^g,x_2^g\right>)=\frac{1}{2}\cdot x_1^g + x_2^g + \frac{1}{2}\cdot(2\cdot t_g-1)\cdot\frac{1}{2}x_1^g + \epsilon^g,
  \end{equation}
где случайные величины $X_1^g$ и $X_2^g$ независимы и взяты из стандартного нормального распределения $N(0,1)$, случайный шок получен из нормального распределения $N(0,0.1)$, а воздействие $t_g \in \{0,1\}$ -- это случайная величина такая, что $\PP{t_g=1}=0.5$. В соответствие с законом~(\ref{eq:synt-data2}), авторы порождали выборки в 1000 элементов и проверяли насколько корректен их подход к расчёту доверительных интервалов, предложенный в этой работе. Другие законы порождения синтетических данных, используемые в различных работах, могут являться более сложными и их подробное рассмотрение выходит за рамки этой главы.

Практически любой новый метод можно проверить, по крайней мере, таким образом. Однако синтетические данные не отражают все богатство реального мира, поэтому также используются и другие подходы к проверке качества того или иного метода.

\subsection{Проверка качества работы подходов на реальных данных}
\addtocru{subsection}{Проверка качества работы подходов на реальных данных} 
\addtocen{subsection}{Model Verification on Real Data} 

Антиподом описанного выше метода проверки качества работы различных алгоритмов, является проверка этих алгоритмов на реальных данных. На этих данных мы не знаем истинных подгрупп или истинных эффектов от воздействия. В этом случае простейшим способом проверки является соотнесение результатов работы метода с известными фактами или с мнением экспертов~\cite{Manahan2005,Athey2016,Taddy2016}. Такая оценка качества работы не является количественной и является достаточно субъективной.

Существуют и более объективные подходы к проверке качества различных методов на реальных данных. Такие подходы часто предполагают разделение выборки данных на обучающую и проверочную\footnote{Обычно в машинном обучении используется термин <<тестовая выборка>>, однако для избежания путаницы между терминами <<тестовая группа>> и <<тестовая выборка>> будем пользоваться термином <<проверочная выборка>>} подвыборки случайным образом. Наблюдения из проверочной подвыборки не передаются для построения модели. При этом после построения модели её качество оценивается только на этой проверочной выборке данных.

Так, например, в \cite{Weisberg2015} использованы данные случайного слепого исследования вещества <<спиронолактон>> для помощи пациентам с сердечной недостаточностью. Пациентов случайном образом разделили на тестовую и контрольную группы. Тестовой группе давали спиронолактон, а контрольной плацебо. Спиронолактон показал свою высокую эффективность в таком исследовании~\cite{Pitt1999}. Однако в последствии стало известно, что в некоторых пациентах спиролактон может вызывать гиперкалиемия (повышенное содержание калия в крови)~\cite{Juurlink2004}. В работе~\cite{Weisberg2015} авторы задались вопросом выделить характеристики пациентов, для которых спиролактон вызывает гиперкалиемию. Для этого на данных случайного слепого исследований спиронолактона, они построили модель, предсказывающую ожидаемое максимальное повышение калия в крови в течении первых 12 недель приёма препарата в зависимости от характеристик пациента (всего 63 характеристики). Данная модель обучалась на 80\% наблюдений слепого случайного исследования препарата спиронолактон. Затем обученная модель для каждого пациента из оставшихся 20\% наблюдений предсказывала на сколько в данном пациенте увеличится уровень калия после приёма тестируемого препарата. Для этих пациентов помимо предсказания изменения уровня калия ($\Delta\hat{Y}_g$), также известны принимал ли он спиронолактон ($t_g=1$) и максимальный уровень калия в течении первых 12 недель $Y_g(t_g)$. На основании этих данных можно проверить, насколько предсказанное изменение калия $\Delta\hat{Y}_g$ соответствует фактическим данным эксперимента $t_g$ и $Y_g(t_g)$.

Для этого авторы построили простую линейную модель (оценивались коэффициенты $\alpha_i$) следующего вида:
\begin{equation}
  \label{eq:spironolactone}
  Y_g = \alpha_0 + \alpha_1 \cdot t_g + \alpha_2 \cdot \Delta\hat{Y}_g + \alpha_3 \cdot t_g \cdot \Delta\hat{Y}_g,
\end{equation}
В результате было обнаружено, что коэффициенты $\alpha_1$ и $\alpha_2$ не были статистически значимо отличны от нуля, в то время как коэффициент $\alpha_3$ был статистически значимо отличен от нуля и положителен, что доказывает, что разработанная в статье модель, действительно даёт предсказания коррелированные с истинным эффектом. Несмотря, на то, что такой подход позволяет оценить работает ли та или другая модель, сравнивать модели между собой этим подходом к проверке моделей затруднительно.

Для сравнения качества работы различных моделей оценки эффекта от воздействия и выделения \SDuTE на реальных данных может служить средний эффект по уровню $\alpha$ и связанный с этим расчётный индекс Qini~\cite{Hansotia2002,Radcliffe2007,RadcliffeSurry2011,Guelman2015a,Guelman2015}. В этом случае необходимо также разбить выборку данных на обучающую и проверочную подвыборки. Далее несколько моделей оценки эффекта от воздействия могут быть обучены на обучающей выборке. В последствии их сравнение должно производиться только с участием проверочной подвыборки данных. Рассмотрим эти методы подробнее.

Пусть есть некоторый метод, умеющий рассчитывать для каждого наблюдения $g$ эффект от воздействия $\Delta\hat{Y}_g$. Тогда зная оценку этого эффекта некоторой моделью для проверочной подвыборки данных, можно упорядочить все наблюдения проверочный подвыборки по убыванию предсказанного эффекта. Далее из проверочной подвыборки данных отбираются только лучшие наблюдения (в пропорции $\alpha$ от все проверочной подвыборки данных) с точки зрения предсказанного эффекта (имеют наибольший эффект). В этом случае $\alpha$ является параметром сравнения, задаваемым исследователем. Далее, на этой части выборки можно оценить истинный средний эффект от воздействия (и его доверительный интервал) и затем уже полученные оценки средних эффектов от воздействия для разных моделей и сравнивать. Интуиция тут следующая: если мы возьмём идеальную модель, то она для наблюдений с большими индивидуальными эффектами от воздействия будет предсказывать большие эффекты, и как следствие в выбранной подвыборке проверочной выборки данных средний эффект должен быть большим. С другой стороны, если модель даёт плохие предсказания и как следствие упорядочивает наблюдение в случайном порядке, то средний эффект от воздействия не будет отличаться от среднего эффекта на всей проверочной выборке. Значит различные модели оценки индивидуального эффекта от воздействия можно сравнивать по размеру среднего эффекта от воздействия (и его доверительного интервала) на подвыборке проверочный выборки данных, <<выбираемой>> ($\alpha$ наблюдений с наибольшим предсказанием модели) той или иной моделью.

В качестве примера рассмотрим таблицу~\ref{tbl:testing-alpha}, в которой представлена проверочная выборка данных, а также результат предсказания индивидуального эффекта от воздействия $\Delta\hat{Y}_g$ для каждого наблюдения проверочной подвыборки данных некоторой моделью. В этом примере выборка данных упорядочена по столбцу $\Delta\hat{Y}_g$, поэтому уникальные идентификаторы в первой колонке перемешаны. Отметим, что в последних двух колонках, находятся результаты случайного слепого исследования, которые не использовались для обучения. Как следствие, эти данные можно использовать для оценки качества модели оценки индивидуального эффекта от воздействия, давшей предсказания эффекта в колонке $\Delta\hat{Y}_g$. Далее, задав долю $\alpha=66.7\%$, наша модель <<выбрала>> бы только лучшие 4 наблюдений (в таблице~\ref{tbl:testing-alpha} эти наблюдения отделены двойной горизонтальной линией). Средний эффект от воздействия на этих наблюдениях равен $\frac{-0.48+ 0.35}{2} - \frac{-0.34-1.01}{2} = 0.91$. Если бы наша модель упорядочивала бы наши наблюдения случайным образом, то средний эффект от воздействия на случайных $66.7\%$ наблюдениях выборки был бы равен среднему эффекту от воздействия на всей выборке $\frac{-0.48+ 0.35-1.12}{3} - \frac{-0.34-1.01+0.99}{3} = -0.3$ (в данном примере мы естественно не рассматриваем доверительные интервалы в силу малости рассматриваемых выборок данных). Таким образом, каждому упорядочиванию наблюдений в проверочной выборке данных (как следствие каждой модели предсказания индивидуального эффекта от воздействия) можно сопоставить число (средний эффект от воздействия на лучших $\alpha$ наблюдениях в соответствии с задаваемым порядком) и его доверительный интервал, которые отражают качество каждой модели оценки индивидуального эффекта от воздействия.

\begin{table}
  \caption{Проверочная выборка данных с предсказанным индивидуальным эффектом от воздействия.}
  \label{tbl:testing-alpha}
  \centering
  \begin{tabular}{l|c|cc}
    \hline\hline
    & 
    $\Delta\hat{Y}_g$& $Y_g(t_g)$ & $t_g$ \\
    \hline
    $g_{6}$ &
     1.42 & -0.34 & 0 \\
    $g_2$ &
     1.02 & -0.48 & 1 \\
    $g_3$ &
     0.66 & 0.35 & 1 \\
    $g_5$ &
     0.49 & -1.01 & 0 \\
    \hline
    \hline
    $g_4$ &
     -0.64 & 0.99 & 0 \\
    $g_1$ &
     -0.9 & -1.12 & 1 \\
    \hline\hline
  \end{tabular}
\end{table}

Таким образом зафиксировав $\alpha$ можно сравнивать различные модели. Вариацией этого подхода является рассмотрение $\alpha$ худших наблюдений проверочной выборки, либо наблюдений соответствующих некоторому квантилю распределения предсказаний $\Delta\hat{Y}_g$. В частности в работах~\cite{Guelman2015a,Guelman2015} различные модели сравниваются на децилях распределения предсказаний $\Delta\hat{Y}_g$.

Если уровень $\alpha$ сложно выбрать для сравнения моделей, то можно поступить следующим образом. Для каждой модели $\mathtt{M}$ и каждого $\alpha$ можно посчитать значение оценки $\tau_{\mathtt{M}}(\alpha)$ среднего эффекта от воздействия на части $\alpha$ проверочной выборки данных, отражающее качество модели $\mathtt{M}$ для соответствующей пропорции $\alpha$.  
Для любого $\alpha$ можно рассмотреть также выбор случайных $\alpha$ наблюдений, получая средний эффект $\tau_{rnd}$, совпадающий со средним эффектом на всей выборке. Тогда, каждую модель будет характеризовать кривая $\tau_{\mathtt{M}}(\alpha) - \tau_{rnd}$, показывающая на сколько лучше модель $\mathtt{M}$ предсказывает индивидуальный эффект от воздействия, чем <<случайная>> модель для пропорции $\alpha$. Эта кривая почти соответствует QINI-кривой с одним уточнением.

В большинстве маркетинговых задач, исследователя интересует не столько средний эффект от воздействия, сколько общий эффект, т.е. средний эффект умноженный на количество клиентов, подвергшихся воздействию. Действительно, если воздействовать только на 1\% клиентов со средним эффектом 100 рублей, это будет менее выгодно, чем воздействовать на 50\% клиентов со средним эффектом 20 рублей: в первом случае общий выигрыш фирмы будет 1 рубль на каждого клиента фирмы, а во втором 10 рублей на каждого клиента фирмы, поэтому QINI кривая должна быть ещё домножена на количество наблюдений под воздействиям, т.е. на долю $\alpha$. Таким образом, Qini-кривая имеет следующий вид $\mathtt{QINI}(\alpha)=\alpha(\tau_{\mathtt{M}}(\alpha) - \tau_{rnd})$, которая соответствует общему эффекту от воздействия на доле $\alpha$ всей проверочной выборки с поправкой на средний эффект от воздействия на всей проверочной выборке данных. 

Тогда Qini-индекс соответствует площади под графиком $\alpha(\tau(\alpha)-\tau_{rnd})$, агрегируя таким образом качество работы той или иной модели в одно число. Это число показывает насколько в среднем лучше работает тестируемая модель, чем подход формирующий \SDuTE случайным образом. 
Например, для нашего примера в таблице~\ref{tbl:testing-alpha}, функция $\mathtt{QINI}(\alpha)$ в точках 0, $\frac{1}{3}$, $\frac{2}{3}$, 1 имела бы следующие значения: 0, $\frac{1}{3}\left(  -0.48 - (-0.34)  - 0.3 \right) = 0.06$, $\frac{2}{3}\left(  0.91 - (-0.3) \right) = 0.81 $, 0. Нужно отметить, что при $\alpha=1$ всегда получается 0, так как вне зависимости от модели (от порядка наблюдений в соответствии с моделью) мы всегда возьмём всю выборку данных, а средний эффект от всей выборки не зависит от порядка и значит качество любой модели на всей выборке равно качеству случайной модели.

Качество многих разработанных моделей проверялось на различных реальных выборках данных в таких областях как медицина~\cite{Cai2010,Lipkovich2011,VanKruijsdijk2015,Weisberg2015,Chekroud2016}, маркетинг и взаимодействие с клиентами~\cite{Hansotia2002,Manahan2005,Guelman2015a,Guelman2015}, интернет-коммерция~\cite{Taddy2016}. Однако, необходимо отметить, что в большинстве работ качество работы различных методов на этих реальных выборках данных не сравнивалось между собой. Наиболее подробное сравнение различных моделей представлено в работах~\cite{Guelman2015} и~\cite{Devriendt2018}




%
\section{Существующие подходы к выделению \SDuTE}
\addtocru{section}{Существующие подходы к выделению \SDuTE} 
\addtocen{section}{Methods of Treatment Effect Estimation} 

В этом разделе будут рассмотрены некоторые методы решения задач (\ref{eq:task1}) и (\ref{eq:task2}). На данный момент есть несколько обзорных работ, систематизирующих существующие работы в этой области~\cite{SotysMichaandJaroszewicz2015,Gutierrez2017,Devriendt2018}. Существующие методы, в первую очередь, отличаются по виду переменой отклика $Y$, которая может быть как численной, так и категориальной (обычно бинарной). Так если нас интересует, сколько дополнительных денег мы заработали маркетинговой кампаний, то $Y$ численная и соответствует тратам клиентов за определённый период после акции. Если целью маркетинговой кампании было привлечь новых клиентов, то в качестве переменной отклика будет регистрироваться лишь факт того, что клиент совершил хотя бы одну покупку.

Другим существенным различием существующих моделей, является методология положенная в их основу. 
Здесь все модели можно разделить на две большие группы. В первую входят модели, которые специальным образом трансформируют задачу оценки индивидуальных эффектов от воздействия в задачу регрессии (если переменная отклика численная) или в задачу классификации (если переменная отклика категориальная). Обе эти задачи можно описать следующим образом. Пусть есть некоторая переменная отклика $Y$ и независимые переменные $X$ как можно найти наиболее точную функциональную зависимость между $Y$ и $X$ ($Y \approx f(x)$). Как можно заметить, в задачах регрессии и классификации отсутствует переменная воздействия $t_g$, и основной вопрос решаемый в таких моделях -- каким именно образом можно закодировать переменную воздействия, чтобы решение задачи классификации или регрессии давало решении задачи оценки индивидуальных эффектов от воздействия или задаче выявления \SDuTE.

Во вторую группу входят специально разработанные модели для решения задач (\ref{eq:task1})  и (\ref{eq:task2}). Большинство таких методов являются моделированными моделям машинного обучения для решения задач классификации и регрессии. В частности многие модели этой группы являются моделями деревьев решений или ансамблями деревьев решений. Далее в этом разделе рассмотрим подробнее модели этих двух групп.

\subsection{Методы сведения к задаче регрессии}
\addtocru{subsection}{Методы сведения к задаче регрессии} 
\addtocen{subsection}{Methods of Reduction to Regression} 

Одними из первых и самых простых подходов к сведению задачи оценки индивидуальных эффектов от воздействия или задачи выделения \SDuTE являются модели основанные на подходе независимого оценивания\footnote{\textit{англ.} indirect estimation methods.}, известного также как подход двух моделей. Этот подход разделяет имеющиеся данные на две части, на каждой из которых строит независимую модель. Альтернативным методом сведения задачи оценки индивидуальных эффектов к задачам регрессии или классификации являются методы, которые специальным образом трансформируют данные, на которых строится только одна модель. 

\subsubsection{Подход независимого оценивания}

Основной идей подхода независимого оценивания является построение двух моделей, предсказывающих значение переменной отклика $Y$ по свойствам каждого описания объекта $d \in D$. Первая модель строится по тестовой подгруппе, то есть по той, которая получила воздействие, а вторая --- по контрольной подгруппе. Если предположить, что эти модели могут достаточно точно предсказывать среднее значение переменной отклика в каждой из ситуаций, то разница в предсказаниях моделей даст нам средний эффект от воздействия для наблюдений с характеристиками $d$. 

В качестве примера, рассмотрим выборку данных представленную в таблице~\ref{tbl:synt-data}. Для того, чтобы применить подход независимого оценивания, нам сначала эту выборку данных нужно разделить на две подвыборки. В первую войдут наблюдения, на которых было оказано тестовое воздействие ($t_g=1$), а во вторую -- контрольное воздействие ($t_g=0$). Так в первую выборку войдут наблюдения $g_1$, $g_2$, и  $g_3$, а во вторую $g_4$, $g_5$ и $g_6$. В каждой из этих подвыборках можно убрать переменную воздействия, так как она не меняется внутри этих подвыборок. Далее на каждой подвыборке мы можем построить свою регрессионную модель, которая будет предсказывать переменную отклика $Y_g(t_g)$. Так по первой выборке мы построим модель $\mathbb{E}[Y \mid x_1,x_2,t_g=1] = \mu_1(x_1,x_2)$, которая оценивает математическое ожидание переменой отклика для наблюдений, на которых было оказано тестовое воздействие. А на второй выборке мы построим модель $\mathbb{E}[Y \mid x_1,x_2,t_g=0]=\mu_2(x_1,x_2)$, оценивающую математическое ожидание переменной отклика, для наблюдений на которых было оказано контрольное воздействия.

Если вспомнить, что данные в таблице~\ref{tbl:synt-data} были построены в соответствии с законом~(\ref{eq:synt-data}), то идеальные модели $\mu_1$ и $\mu_2$ должны иметь следующий вид $\mu_1(x_1,x_2)= x_1 + x_2$, а $\mu_2(x_1,x_2) = x_2$. Как только мы построили две независимые модели на тестовой и контрольной подвыборках, мы можем оценить индивидуальный эффект от воздействия $\Delta Y_g = \mu_1(x_1^g,x_2^g) - \mu_0(x_1^g,x_2^g)$ для любого наблюдения $g$ и его описания $\delta(g)=\left<x_1^g,x_2^g\right>$.

Разделив обучающую выборку на две подвыборки, можно обучать две любые регрессионные модели. Это могут быть как простые линейные модели, так и современные модели машинного обучения, такие как градиентный бустинг~\cite{Friedman2001}. Например, в \cite{Hansotia2002,Hansotia2002a} эффект от воздействия предсказывается как разница двух логистических регрессий и как разница двух деревьев решений. В частности, Хансотия и Руксталес\footnote{\textit{англ.} Hansotia \& Rukstales} показывают, что деревья решений лучше подходят для задачи предсказания эффекта от воздействия, чем логистическая регрессия. В \cite{Manahan2005} также предлагается подход независимого оценивания. Автор предлагает оценивать эффект от воздействия как разницу предсказания нейронной сетью переменой отклика для тестовой и контрольной групп. Автор показывает, что нейронная сеть даёт лучший результат, чем логистическая регрессия.

Ещё один метод, близкий по своей сути к методам независимого оценивания является метод $K$-ближайших соседей~\cite{Alemi2009}. Для каждого нового наблюдения авторы предлагают упорядочивать все наблюдения обучающей выборки с точки зрения расстояния до него от нового наблюдения. Затем авторы берут такое минимальное количество ближайших наблюдений из обучающей выборки, что среди этих наблюдений есть статистически значимое различие между тестовой и контрольной группой. Если бы в такой работе использовался стандартный метод $K$-ближайших соседей для фиксированного $K$ для тестовой и контрольной подвыборок, то был бы стандартный подход двух моделей. В данном случае, для каждого предсказываемого наблюдения $g$ значение $K$ выбирается автоматически так, чтобы быть уверенным в статистически значимом различии между тестовой и контрольной подвыборками среди  $K$ ближайших соседей.

Здесь стоит отметить, что существует большое количество прикладных работ~\cite{Brun2002,Hansotia2002,Cai2010,Dorresteijn2011,Imai2013,VanderLeeuw2014,VanKruijsdijk2015,Chekroud2016}, которые не всегда говорят напрямую про эффект от воздействия, и решают задачу классификации или регрессии в определённой предметной области. Однако авторов этих работ интересует именно оценка индивидуального эффект от воздействия, а не предсказание ожидаемого значения переменной отклика $Y_g$. В самих работах, после обучения нескольких моделей классификации или регрессии, расчитывается индивидуальный эффект от воздействия как разница обученных моделей и, таким образом, работы опираются на вариацию подхода независимого оценивания. Так, например, в~\cite{VanKruijsdijk2015} исследуется взаимосвязь между долгосрочным приёмом аспирина и риском раковых заболеваний, сердечно-сосудистых заболеваний и желудочно-кишечных кровотечений. Для этого на большой выборке данных пациентов, получающих регулярно аспирин или плацебо, строится несколько моделей пропорциональных конкурирующих рисков\footnote{
  A Proportional Hazards Model for the Subdistribution of  a Competing Risk~\cite{Fine1999}.
} для каждого из видов заболеваний. При этом отдельно строится модель для пациентов получающих аспирин и отдельно для пациентов, получающих плацебо. В итоге эффект от аспирина для отдельного клиента предсказывался как разница предсказний для этого пациента согласно моделям пациентов, получающих аспирин, и пациентов, не получающих аспирин.

В целом, методы независимого оценивания могут быть основаны на любом существующем методе решения задач классификации или регрессии. Однако на практике такие методы работают не очень хорошо, что, вероятно, связано с тем, что каждая модель оценивается лишь на половине выборки данных, а также на том, что ошибки каждой из моделей приводят к увеличению ошибки оценки разности этих двух моделей. Более точным подходом является специальное преобразование входных данных таким образом, чтобы получить задачу регрессии и решать её существующими стандартными методами. 

\subsubsection{Подходы модификации данных}

%
%

Одним из простейших методов модификации данных для сведения задачи оценки индивидуальных эффектов от воздействия к задачи регрессии или классификации является добавление специальных переменных взаимодействия\footnote{\textit{англ.} interaction term}, которые на контрольной группе принимают значение ноль, а на тестовой соответствуют независимым переменным исходной выборки данных~\cite{Lo2002,Selker1997,Kent2002,Dorresteijn2013,Burke2014,Krintel2015}. В этом случае обучается модель вида $\hat{Y}=f(X,T,TX)$\footnote{Напомним, что большие заглавные буквы обозначают случайные величины, в частности $T$ -- это случайная величина соответствующая назначению воздействия объектам.}. Такого рода модели хорошо изучены в математической статистике и эконометрике, и, соответственно, существуют строгие методики расчёта статистической значимости и доверительных интервалов для многих спецификаций функции $f(\cdot)$, что является несомненным достоинством этого подхода. Обучив такую модель, эффект от воздействия может быть оценен как 
\begin{equation}
  \label{eq:predict-interaction}
  \hat{\Delta Y}_g=\hat{\tau}(\delta_g) = \hat{\tau}(X)=f(X,1,X) - f(X, 0, 0 \cdot X). 
\end{equation}
Рассмотрим данные представленные в таблице~\ref{tbl:synt-data}. Тогда добавление переменных взаимодействия, трансформирует выборку данных в данные показанные в таблице~\ref{tbl:data-interaction}. Как мы видим значение последних двух переменных в трансформированной выборке данных для контрольной подвыборке всегда равно нулю. Данные в таблице~\ref{tbl:synt-data} были созданы в соответствии с законом~(\ref{eq:synt-data}). Следовательно. идеальная модель регрессии для этой выборки данных должна иметь следующий вид:
\begin{equation*}
  \hat{Y}(t_g,x_1^g,x_2^g,t_gx_1^g,t_gx_2^g) = 0 + 0\cdot t_g + 0 \cdot x_1^g + x_2^g + t_gx_1^g + 0 \cdot t_gx_2^g 
\end{equation*}

\begin{table}
  \caption{Трансформированная путём добавления переменных взаимодействия выборка данных из таблицы~\ref{tbl:synt-data}.}
  \label{tbl:data-interaction}
  \centering
  \begin{tabular}{l||c|c|cc|cc}
    \hline
    \hline
    &
    $Y_g$ & $t_g$ & $x_1^{(g)}$ & $x_2^{(g)}$  & $t_g \cdot x_1^{(g)}$ & $t_g \cdot x_2^{(g)}$ \\
    \hline
    $g_1$ &
    -1.12 & 1 & -0.9 & -1.56 & -0.9 & -1.56 \\
    $g_2$ &
    -0.48 & 1 & 1.02 & -1.07 & 1.02 & -1.07 \\
    $g_3$ &
    0.35 & 1 & 0.66 & -0.14 & 0.66 & -0.14 \\
    $g_4$ &
    0.99 & 0 & -0.64 & 0.1 & 0 & 0 \\
    $g_5$ &
    -1.01 & 0 & 0.49 & -0.41 & 0 & 0 \\
    $g_{6}$ &
    -0.34 & 0 & 1.42 & 0.38 &  0 & 0 \\
    \hline
    \hline
  \end{tabular}
\end{table}
Однако уже из формулы~(\ref{eq:predict-interaction}) предсказания эффекта от воздействия видно, что ошибка предсказания будет состоять из двух ошибок: для предсказания $Y$ при наличии воздействия, и для предсказания $Y$ без воздействия. Поэтому методы этого подхода безусловно напоминают методы подхода двух моделей. Более того, как правило, каждой из переменных $X,T$ и $X\cdot T$ соответствует отдельный коэффициент в модели $f$, что приводит к тому, что коэффициенты соответствующие переменным $X \cdot T$ оцениваются лишь на части выборки данных, и соответсвенно к большей ошибке в оценке этих коэффициентов. Однако именно эти коэффициенты представляют наибольшую важность при прогнозировании эффекта от воздействия.


Для того, чтобы решить эти проблемы, было предложено преобразовать переменную $Y$ таким образом, чтобы решением задачи регрессии на преобразованных данных был эффект от воздействия~\cite{Jaskowski2012,Weisberg2014,Weisberg2015}. 
Пусть сначала целевая переменная $Y$ -- это бинарная переменная принимающая значения 0 и 1, где 1 считается хорошим исходом, а воздействие может изменить вероятность того, что $Y$ примет одно из этих двух значений. Так, в~\cite{Jaskowski2012,Weisberg2014} предлагается ввести следующую случайную величину: $Z=Y\cdot T + (1-Y)\cdot(1-T)$, то есть для конкретного наблюдения $g$, $z_g$ принимает значение 1, если получен хороший исход ($y_g=1$) и при этом было воздействие ($t_g=1$) или, если получен негативный исход и воздействие не оказывалось. Если размер тестовой и контрольной группы равны, то есть $\mathbb{P}(T=1)=0.5$, тогда не сложно видеть, что 
\begin{equation*}
  \EE{Z}=\EE{Y\mid T = 1}\cdot\PP{T=1} + \EE{1-Y\mid T=0}\cdot\PP{T=0} = \frac{1}{2}\cdot(1 + \EE{\Delta Y}),
\end{equation*}
то есть монотонно зависит от эффекта воздействия $\Delta Y$. Тогда, при оценке модели $Z=f(X)+\epsilon$, полученная модель $f(X)$ будет предсказывать $\EE{Z \mid X}$,  что также будет являться монотонным преобразованием от воздействия.

Вайсберг и Понт\footnote{\textit{англ.} Weisberg \& Pontes} на основе подхода, рассмотренного выше, опубликовали метод работы с численной переменной~$Y$~\cite{Weisberg2015}. Если $Y$ -- действительная переменная отклика, то можно ввести случайную величину $Z=2\cdot Y\cdot(2\cdot T -1)$, т.е. для наблюдения $g$, реализация случайной величины $z_g$ равна переменной отклика $у_g$, если было оказано воздействие $t_g=1$, а если $t_g=0$, то $z_g=-y_g$. Математическое ожидание такой случайной величины $Z$ равно математическому ожиданию эффекта от воздействия $\EE{Z}=\EE{\tau}$. Аналогично, оценка модели $Z=f(X) + \epsilon$ приводит к получению функции $f(X)$, которая предсказывает $\EE{Z\mid X}=\tau(X)$.

Вернёмся к данным из таблицы~\ref{tbl:synt-data}. Преобразованные данные показаны в таблице~\ref{tbl:weisberg-data}. В этих данных уже отсутствует переменная $y_g$, которая была заменена на переменную $z_g$. Также в трансформаированных данных отсутствует переменная отклика $t_g$. Так как данные в таблице~\ref{tbl:synt-data} были созданы в соответствии с законом~(\ref{eq:synt-data}). Следовательно. идеальная модель регрессии для трансформированной выборки данных их таблицы~\ref{tbl:weisberg-data} должна иметь следующий вид:
\begin{equation*}
  \hat{Z}(x_1^g,x_2^g) = 0 + x_1^g,
\end{equation*}
т.е. модель имеет в точности необходимый вид для предсказания эффекта от воздействия. В данном случае для обучения такой модели уже используется вся выборка данных, и как следствие качество предсказания такой модели должно быть лучше всех рассмотренных ранее моделей.

\begin{table}
  \caption{Трансформированная путём трансформации переменной отклика $y_g$ выборка данных из таблицы~\ref{tbl:synt-data}.}
  \label{tbl:weisberg-data}
  \centering
  \begin{tabular}{l||c|cc}
    \hline
    \hline
    &
    $z_g$ &  $x_1^{(g)}$ & $x_2^{(g)}$ \\
    \hline
    $g_1$ &
    -1.12 & -0.9 & -1.56 \\
    $g_2$ &
    -0.48 & 1.02 & -1.07 \\
    $g_3$ &
    0.35 & 0.66 & -0.14 \\
    $g_4$ &
    -0.99 & -0.64 & 0.1 \\
    $g_5$ &
    1.01 & 0.49 & -0.41 \\
    $g_{6}$ &
    0.34 & 1.42 & 0.38 \\
    \hline
    \hline
  \end{tabular}
\end{table}


В некотором роде обобщением таких работ является работа Тианом с соавторами\footnote{\textit{англ.} Tian et al.}~\cite{Tian2014}. Действительно, если есть возможность изменить целевую переменную $Y$, то аналогичное преобразование может быть применено и к независимым переменным $X_i$. 
Действительно, рассмотрим линейную модель $Z = Y \cdot (2\cdot T - 1) = \alpha \cdot X + \epsilon$. Умножим левую и правую часть на $2\cdot T - 1$, принимающего значения из $\{-1,1\}$, получим модель $Y = \alpha \cdot \left(X \cdot (2\cdot T - 1)\right) + \epsilon$. В этой модели вместо преобразования переменной отклика, трансформируются независимые переменные $X_i$ как $X_i^*=(2 \cdot T - 1) \cdot X_i$.
Такое преобразование будет эквивалентным предыдущим подходам в случае линейной модели. Однако данное преобразование независимых переменных также позволяет оценивать логит-модель\footnote{Случай когда переменная отклика $Y$ является бинарной.} и некоторые другие типы моделей\footnote{Подробное описание этих моделей выходит за рамки этой главы. Заинтересованный читатель может обратиться к работе~\cite{Tian2014}}.
Здесь нужно отметить, что для каждого из этих видов моделей, авторы формально доказывают, что оценённая модель по их методологии будет наилучшей даже в том случае, когда используемая модель неверно специфицирована, т.е. когда, например, при квадратичной зависимости между $X_i$ и эффектом от воздействия $\tau$ обучалась линейная модель.

Ещё раз рассмотрим выборку данных из таблицы~\ref{tbl:synt-data}. Преобразованные данные по методу Тиана и его соавторов показаны в таблице~\ref{tbl:tian-data}. В этом случае идеальная модель, соответствующая закону (\ref{eq:synt-data}) порождения данных, будем иметь в точности такой же вид как и в предыдущем случае:
\begin{equation*}
  \hat{Y}(x_1^g,x_2^g) = 0 + x_1^g,
\end{equation*}
т.е. будет напрямую предсказывать эффект от воздействия для наблюдения $g$. Здесь следует подчеркнуть, что для предсказания эффекта от воздействия нужно передавать непреобразованные независимые переменные, что видно и по форме модели.

\begin{table}
  \caption{Трансформированная путём трансформации независимых переменных $x_g$ выборка данных из таблицы~\ref{tbl:synt-data}.}
  \label{tbl:tian-data}
  \centering
  \begin{tabular}{l||c|cc}
    \hline
    \hline
    &
    $y_g$ &  $x_1^{*(g)}$ & $x_2^{*(g)}$ \\
    \hline
    $g_1$ &
    -1.12 & -0.9 & -1.56 \\
    $g_2$ &
    -0.48 & 1.02 & -1.07 \\
    $g_3$ &
    0.35 & 0.66 & -0.14 \\
    $g_4$ &
    0.99 & 0.64 & -0.1 \\
    $g_5$ &
    -1.01 & -0.49 & 0.41 \\
    $g_{6}$ &
    -0.34 & -1.42 & -0.38 \\
    \hline
    \hline
  \end{tabular}
\end{table}

Большим преимуществом модели Тиана является её простота и возможность использования в сочетании со стандартным статистическим аппаратом для расчёта статистической значимости и доверительных интервалов. Тут нужно отметить, что такой подход должен работать хорошо не только с классическими статистическими моделями, но и с различными методами машинного обучения в силу формальных результатов полученных Тианом и соавторами.

Отметим также, что этот метод сразу оценивает эффект от воздействия, а коэффициенты модели оцениваются на всём множестве наблюдений в отличие от моделей с взаимодействием воздействия и переменных.

Стоит отметить, что рассмотренные здесь модели можно также применять и в случаях различного размера тестовой и контрольной групп. Для этого необходимо взвесить наблюдения в контрольной и тестовой группах таким образом, чтобы суммарный вес наблюдений в этих подгруппах был равным. При оценке самих моделей такое взвешивание лишь меняет функцию потерь, например, средне-квадратичное отклонение, и оно корректно работает с большинством из существующих методов.

Также необходимо отметить ряд недостатков методов сведения задачи выявления \SDuTE к задаче регрессии (или классификации). Прежде всего на практике такие методы применялись лишь в случае обобщённых линейных моделей\footnote{\textit{англ.} generalized linear models}, и, соответственно, такие модели сложно использовать при наличии нелинейной зависимости между независимыми переменными и эффектом от воздействия.

Вторым недостатком таких методов является их зависимость от соотношения числа наблюдений в контрольной и тестовой группах. И если на больших подвыборках исходных данных качество оценки эффекта от воздействия вероятно будет оцениваться достаточно точно, то чем меньшую подгруппу потребуется выделять, тем больше вероятность некорректной оценки эффекта от воздействия на этой выборки, так как соотношение между числом наблюдений в контрольной и тестовой подгруппах вероятно существенно изменится.

Все большее число работ исправляет эти недостатки посредством адаптации зарекомендовавших себя методов машинного обучения для решения задачи выявления эффекта от воздействия.

\subsection{Адаптированные методы машинного обучения}
\addtocru{subsection}{Адаптированные методы машинного обучения} 
\addtocen{subsection}{Adopted Methods from Machine Learning} 

Основная цель методов машинного обучения -- нахождение моделей, дающих хорошее предсказание. Соответственно линейные модели составляют лишь малую часть всех существующих методов машинного обучения. Нелинейные модели позволяют найти сложную разделяющую поверхность между классами, или сложную зависимость между зависимой и независимыми переменными. Часто такие нелинейные модели дают высокое качество предсказания.

Одной из старейших нелинейных моделей является дерево решений. Соответственно множество работ в области выявления \SDuTE пытаются адаптировать эту модель для решенная задач \ref{eq:task1} и \ref{eq:task2}. Построение дерева решений по данным является рекурсивной процедурой~\cite{Breiman1984}. На каждом шаге есть некоторая часть исходной выборки данных, и необходимо принять решения, нужно ли данную часть выборки данных делить дальше на подвыборки, и если нужно, то как составить предикат (вопрос к наблюдениям), разделяющий рассматриваемую часть данных на подвыборки. Как правило, составление такого предиката разбивается на выбор наилучшей независимой переменной, по которой будет проходить разбиение, и на выбор самого порога разбиения, по которому наблюдения разделятся на две части: те наблюдения, которые имеют значение выбранной переменой выше порога, и те, которые ниже. 

В силу того, что процедура построения дерева является рекурсивной, и на каждом шаге подвыборка исходных данных разбивается на ещё меньши подвыборки, то возникает вопрос остановки этой рекурсивной процедуры. Как правило такой вопрос решается путём фиксации минимального размера подвыборки, которая может быть получена в результате такого рекурсивного деления. В любом случае подвыборку состоящую из одного наблюдения дальше делить невозможно. Фиксация минимального размера подвыборки (листа) в дереве является искусственым параметром и, как правило, применяется как промежуточный этап в построении дерева. В последствии многие алгоритмы построения дерева решений имеют этап усечение ветвей\footnote{\textit{англ.} pruning}, на котром излишнии разбиения удаляются. Для того, чтобы понять какие разбиения были лишними, для этого этапа используется модифицированная функция оценки качества дерева, в которой помимо качества предсказания обучающей выборки построенным деревом также учитывается и размер самого дерева, т.е. количество разбиений исходной выборки на подгруппы. Ниже мы конкретизируем эти этапы именно для задачи выделения \SDuTE.

Стоит также отметить, что деревья решений позволяют определять важность независимых переменных с точки зрения их влияния на качество предсказания, что необходимо для построения хорошей модели выделения \SDuTE~\cite{Larsen2009,RadcliffeSurry2011,Chekroud2016}.

Одними из первых модифицированных моделей дерева решений были работы~\cite{Radcliffe1999,Chickering2000}. Авторы \cite{Chickering2000} предлагают модифицировать критерий разбиения таким образом, чтобы виртуально делить листовую вершину в дереве по признаку воздействия $T$. Другими словами, алгоритм построения такого дерева решений перебирает все независимые переменные $x \in X$ и все возможные разбиения $ x \leq \theta$. Для каждого конкретного разбиения получающиеся листья дополнительно делятся на контрольную и тестовую группу и считается предсказательная сила полученного дерева. Финальное разбиение выбирается как разбиение максимизирующее предсказательную силу дерева. 

Чтобы продемонстрировать этот подход рассмотрим данные из таблицы~\ref{tbl:synt-data}. Пусть на текущем шаге рекурсивной процедуры построения дерева необходимо определить как разбить эту выборку данных на подвыборки. Для этого у нас есть две переменных $x_1$ и $x_2$, по которым можно делить выборку данных на две в зависимости от выполнения каждым наблюдений условий вида $x_1 < 1.1$  или $x_2 < -0.7$. Далее процедура в простейшем случае перебирает все возможные правила (с точностью до положения границы отсечения между двумя соседними точками), и выбирает оптимальный с точки зрения некоторого критерия. Пусть для примера мы рассматриваем правило $X_1 < 0$ в качестве потенциального разбиения имеющейся выборки данных на две. Это разобьёт выборку данных на две части из 2 и 4 наблюдений, каждая из которых будет дополнительно разбита на тестовую и контрольную подгруппы в соответствии с переменной воздействия $t_g$. Таким образом получиться 4 подвыборки: $\{g_1\}$, $\{g_4\}$, $\{g_2,g_3\}$, $\{g_5,g_6\}$. Предсказательная сила такого дерева с точки зрения среднего квадрата ошибки в листьях дерева равна 0.195 (в первых двух подвыборках ошибка предсказания $Y_g$ равна нулю, во третьей равно 0.17, а в четвертой -- 0.22). Перебрав все возможные разбиения исходной выборки на две можно найти оптимальное разбиение с точки зрения предсказательной силы получаемого дерева.

Здесь нужно отметить, что в таком подходе не учитываются ни вариативность предсказания внутри листьев дерева, ни разница в размерах контрольной и тестовой группах, что приводит к переобучению.
Данная проблемы была частично решена в~\cite{Radcliffe1999}, однако в самой работе не приводился конкретный критерий разбиения вершины дерева. Позже авторы специфицировали этот критерий в~\cite{RadcliffeSurry2011}. В частности, вместо предсказательной силы соответствующей части дерева после разбиения, критерий, рассматриваемый в~\cite{Radcliffe1999}, был основан на разнице в среднем значений переменной отклика $Y$ в тестовой и контрольной подгруппах. Однако сама расчётная разница штрафовалась за существенную разницу в размерах тестовой и контрольной групп. В~\cite{RadcliffeSurry2011} утверждается, что критерии такого плана дают нестабильное качество моделей на разных выборках данных.

В работах \cite{Su2009,RadcliffeSurry2011} также предлагается использовать критерий разбиения вершины дерева на основании статистической значимости в различии распределений переменной отклика $Y$ в тестовой и контрольной подгруппах. Это происходит следующим образом. Для каждого потенциального разбиения строится линейная модель, включающая в себя специальную переменную взаимодействия между воздействием $T$ и тем фактом куда отнесено (в левую или в правую ветвь) каждое наблюдение в дереве. Статистическая значимость коэффициента перед этой переменной и является критерием разбиения. Также нужно отметить, что авторы~\cite{RadcliffeSurry2011} предлагают для усечения деревьев порождать методом бутстрэпа\footnote{Из имеющейся выборки случайно с повторениями вытягивать наблюдения для формирования выборки данных такого же размера как и исходная выборка} несколько случайных выборок данных. На одной из выборок данных происходит обучение, т.е. строится дерево решений, а остальные порождённые выборки данных используются для оценки вариативности предсказания обученного дерева решений. Если вариативность какой-либо вершины превышает некоторый порог, то соответствующее разбиение убирается из дерева решений.

Ещё один подход к построению деревьев решений был предложен в~\cite{Rzepakowski2010,Rzepakowski2012}. Этот подход основывается на критерии разбиения вершины дерева, включающем в себя расстояние между эмпирической плотностью распределения переменной отклика в тестовой и контрольной подгруппах этой вершины дерева. В качестве примеров такого расстояния авторы рассматривают дивергенцию Кулбака-Лейблера и Эвклидово расстояние. Тогда качество того или иного разбиения характеризует насколько сильно увеличилось расстояние между тестовой и контрольной подвыборками после деления этой выборки на две части. Более формально такой критерий написан ниже. Здесь мы предполагаем, что нужно проверить качество некоторого фиксированного разбиения выборки данных, соответствующей текущей вершине дерева. Так, например, выше мы рассматривали правило разбиения $x_1 < 0$. При фиксации такого правила выборка данных делится на две части: левую и правую подветвь дерева. Далее для расчёта критерия качества $Q(\text{split})$ необходимо померить расстояние между двумя распределениями (распределение переменной отклика $Y$ в тестовой и контрольной подгруппах) в каждой из получившихся подветвей дерева. Далее эти расстояния сравниваются с расстоянием тестовой и контрольной выборкой в исходной вершине:
\begin{align*}
  Q(\text{split}) = &\frac{N_\text{Left}}{N_{Left} + N_{Right}}D[Y(T=1 \mid \text{Left}],~Y(T=0 \mid \text{Left})] + \\
  +&\frac{N_\text{Right}}{N_{Left} + N_{Right}}D[Y(T=1 \mid \text{Right}),~Y(T=0 \mid \text{Right})] - \\
  -&D[Y(T=1),~Y(T=0)],
\end{align*}
где $N_\text{Left}$, $N_\text{Right}$ -- размеры всей выборки данных в левой и правой подвершине, а $D[Y_1,~Y_2]$ -- функция определения расстояния между распределениями случайных величин $Y_1$, и $Y_2$.

Вернёмся к выборке данных из таблицы~\ref{tbl:synt-data}. Пусть для простоты распределение случайной величины $Y$ будет характеризоваться вероятностью $p$, что случайная величина $Y$ меньше нуля. Тогда расстояние между распределениями случайных величин $Y_1$ и $Y_2$ будем считать как разницу между $p_1=\PP{Y_1 < 0}$ и $p_2=\PP{Y_2<0}$, т.е. $D[Y_1,~Y_2]=|p_1-p_2|$. Посчитаем, какое расстояние между распределениями переменной отклика $Y$ в тестовой и контрольной подвыборках: в каждой подвыборке два наблюдения из трёх отрицательны, значит расстояние между этими распределениями равно нулю. Разобьем теперь исходную выборку данных на две в соответствии с правилом $x_1 < 0$, тогда в левую подвершину попадут наблюдения $g_1$ (тестовая выборка) и $g_4$ (контрольная выборка), а остальные попадут в правую. В левой подвершине наблюдение из тестовой выборки имеет отрицательную переменную отклика $Y$, а наблюдение из контрольной выборки имеет положительную переменную отклика, значит расстояние между этими распределениями равно 1 ($p_{\text{тест}} = 1$), в правой ветке дерева будет 4 наблюдения, при этом в тестовой выборки переменная отклика принимает одно положительное и одно отрицательное значение, а в тестовой все значения переменной отклика $Y$ отрицательны, значит расстояние между этими распределениями равно $p_{\text{контроль}}=|1-0.5|=0.5$. Таким образом, качество такого разбиения $Q(\text{split})=\frac{2}{6}\cdot 1 + \frac{4}{6}\cdot 0.5 - 0 = 0.67$.

Этот подход также использовался в работах Гуельмана\footnote{\textit{англ.} Guelman} с соавторами для построения случайного леса~\cite{Guelman2012,Guelman2015a}. Позже Гуельман с соавторами усовершенствовали это метод построения дерева решений~\cite{Guelman2015}. Они предложили разделить критерий разбиения вершины дерева на два этапа. На первом этапе они выбирают независимую переменную, а на втором этапе выбирают порог, по которому необходимо разбить эту переменную.

Первый этап этой процедуры построения деревьев основывается на тестах перестановок~\cite{Strasser1999}. 
Общая идея таких тестов в том, чтобы упорядочить все наблюдения в соответствии с рассматриваемой независимой переменой $x_i$. В дальнейшем проверяется, насколько вероятен был наблюдаемый порядок по переменной отклика в предположении, что упорядочивание по переменной $x_i$ не связано с эффектом от воздействия.
Далее для выбора оптимальной точки разбиения используется такой же критерий как и в рассмотренных ранее работах~\cite{Su2009,RadcliffeSurry2011}.

Ещё один метод построения дерева решений для предсказания эффекта от воздействия был предложен Атей и Имбенс\footnote{\textit{англ.} Athey \& Imbens}~\cite{Athey2015,Athey2016}. Их метод является адаптацией методологии построения деревьев классификации и регрессии (CART), однако они напрямую решают задачу, заданную уравнением (\ref{eq:task1}). В работе предполагается, что есть всего два варианта: 1 -- есть воздействия и 0 ($\bot$) -- нет воздействия. 

Отличительной чертой этого подхода является использование так называемого честного оценивания\footnote{\textit{англ.} honest estimation}, заключающегося в том, что выборка объектов разбивается случайным образом на две части: на первой части происходит выделение подгрупп, а на второй части оценивается значение эффекта от воздействия внутри каждой группы. Это позволяет уменьшить эффект переобучения и более точно предсказывать эффект от воздействия.
%
%
Также в этой работе используется особый критерий разбиения вершины дерева на подвершины. Это критерий состоит из двух частей, первая из которых отвечает за ошибку предсказания эффекта от воздействия, а вторая штрафует за малый размер получаемых после разбиения подвыборок. 

В~\cite{Lipkovich2011} авторы вместо построения дерева решений, в котором в каждой вершине происходит одно единственное разбиение, предлагают делить каждую вершину несколько раз с выбором возможно нескольких лучших разбиений. Процесс начинается с одной большой подгруппы, включающей все наблюдения. На каждом шаге поддерживается не более чем $L$ возможно пересекающихся подгрупп наблюдений. Пусть есть некоторая подгруппа, тогда среди переменных, которые ещё не рассматривались для определения этой подгруппы (в самом начале все доступные независимые переменные), и всех возможных разбиений выбираются $M$ лучших (с точки зрения критерия разбиения) разбиений, а из каждого разбиения рассматривается только лучшая подгруппа. 
Сам критерий разбиений основан на расчёте T-статистики на разницу в средних между тестовой и контрольной подвыборками.


У этого метода есть две особенности, которые нужно упомянуть. Для уменьшения случаев переобучения, авторы предлагают корректировать значение \textpi-value, путём проведения случайного эксперимента. В частности, они перемешивают пары $\left<y_g,t_g\right>$, таким образом, что средний эффект от воздействия не меняется, но при этом можно быть уверенным, что нет взаимодействия между независимыми переменными $x_i$ и эффектом от воздействия $\tau$. Тогда, получив значения нужной статистики на этих перемешанных данных можно оценить скорректированное значения \textpi-value для каждой подгруппы.

Вторая особенность метода заключается в том, что он позволяет при очередном разбиении одной из подгрупп, определить критерий разбиения таким образом, что только одна из получаемых подгрупп будет иметь улучшение с точки зрения эффекта от воздействия. С одной стороны, это позволяет выделить подгруппы, которые наиболее сильно привязаны к воздействию, но, с другой стороны, добавляет рисков переобучения. Однако на практике этот метод не всегда применим, так как может обнаруживать только очень сильную неоднородность в эффекте от воздействия. Однако сам подход основывается на достаточно прогрессивных идеях, которые во-видимому можно перенести на другие подходы.

Отдельно упомянем о работе~\cite{Zeileis2008}, которая хоть и не ставит своей целью предсказывать эффект от воздействия, но позволяет строить дерево решений, в листьях которого, можно разместить произвольную модель. В частности, можно разместить простую линейную модель, которая будет предсказывать эффект от воздействия, и при этом всё дерево решений будет иметь сложную нелинейную форму. Этот подход работает следующим образом. Он вписывает модель в рассматриваемом узле дерева и проверяет есть ли <<неслучайность>> в ошибках вписывания модели при упорядочивании этих ошибок по одной из переменных. На текущем шаге построения дерева разбивается та переменная, где статистическая значимость <<неслучайности>> максимальна. Данный метод не нуждается в усечении ветвей, благодаря статистической проверки на неслучайность распределения остатков вписываемой модели. Если неслучайность не может быть обнаружена с достаточным уровнем значимости, то разбиения не происходит и процесс останавливается.

\subsection{Ансамблевые методы}
\addtocru{subsection}{Ансамблевые методы} 
\addtocen{subsection}{Ensemble Methods} 

В предыдущем разделе рассмотрены некоторые методы построения деревьев решений для оценки эффекта от воздействия. Большинство из этих методов использует множественные явные и неявные статистические тесты. Более того, провести коррекцию на множественность тестов зачастую крайне затруднительно, поэтому эти методы склонны к переобучению. Одним из способов борьбы с этой проблемой является создание ансамбля простых классификаторов, которые за счёт некоторой случайности в процессе обучения каждого отдельного элемента ансамбля <<переобучаются>> по разному, что позволяет получить более надёжную модель (ансамбль) как взвешенное голосование её составных частей. 
В нескольких работах показывается, что как и в случае задач классификации и регрессии, применение ансамблевых методов может существенно улучшить качество выделения \SDuTE (или оценки эффекта от воздействия)~\cite{SotysMichaandJaroszewicz2015}. Большая часть ансамблевых методов являются спецификацией или бэггинга~\cite{Breiman1984}, частным случаев которого являются случайные леса, или бустинга~\cite{Freund1997}. Бэггинг -- это ансамбль независимых между собой простых моделей, в то время как бустинг, это последовательный метод обучения простых моделей, где каждая следующая модель <<фокусируется>>  на ошибках предыдущих моделей. Также ансамблевые модели отличаются подходами к получению случайных подвыборок исходных данных, что может происходить как по наблюдениям, так и по независимым переменным~\cite{Fan2003,Geurts2006}. Более того, порождение случайных выборок данных для обучения каждой простой модели может осуществляться на каждом шаге построения простых моделей. 

Гуелман\footnote{\textit{англ.} Guelman} и соавторы предлагают другой ансамблевый подход, называемый Uplift Random Forest~\cite{Guelman2015a}, который адаптирует алгоритм Random Forest~\cite{Breiman2001} для случая выявления \SDuTE. Этот подход работает с бинарными целевыми переменными и бинарными воздействиями, которые часто имеют место в маркетинге. Каждое отдельное дерево строится следующим образом. Сначала случайным образом выбирается часть доступных наблюдений и часть доступных независимых переменных. По такой случайной подвыборке данных строится дерево решений, описанным выше способом, основанном на расчёте расстояний между двумя распределениями. Таким образом получается ансамбль деревьев, каждое из которых может предсказать эффект от воздействия. Результирующее предсказание получается путём усреднения всех предсказаний, входящих в ансамбль.


Авторы усовершенствовали свой подход, чтобы устранить смещённость оценки эффекта от воздействия~\cite{Guelman2015}, основываясь на идеях~\cite{Hothorn2006}. В частности они изменили порядок выбора переменной для разбиения с последующим выбором точки разбиения.
При этом они сначала проверяют статистическую значимость зависимости между переменной отклика и $n$ случайно выбранными зависимыми переменными. 
При отсутствии статистически значимой зависимости, дальнейшее разбиение вершины дерева не производится. Выделив независимую переменную, точка разбиения выбирается путём максимизации взаимодействия между эффектом от воздействия и бинаризацией в этой точке разбиения с точки зрения критерия $\chi^2$.


Несколько работ также формируют ансамбли деревьев на основании своих подходов в области построения деревьев решений для задачи выявления \SDuTE~\cite{RadcliffeSurry2011,Wager2017}, предсказание в большинстве таких работ также получается путём усреднения предсказания каждого отдельного дерева входящего в ансамбль. Отдельно нужно выделить работу~\cite{Wager2017}, в которой авторы также порождают ансамбль простых моделей. Однако здесь авторы идут дальше и разрабатывают теорию, которая позволяет помимо оценки эффекта от воздействия для наблюдения $g$ получать доверительные интервалы этой оценки. Это является несомненным достоинства данного метода, так как позволяет принимать более осознанные решения на основе его предсказаний. 

Принципиально другим подходом к построению деревьев решений является подход Тадди\footnote{\textit{англ.} Taddy} и соавторов. Они предлагают обучать так называемый Байесовский лес для оценки эффекта от воздействия~\cite{Taddy2016}. Этот подход полагается на ансамбль (множество) пар деревьев решений. Одно дерево строится на тестовой ($t_g=1$) подгруппе наблюдений и предсказывает значение целевой переменной $Y_g(t_g)$, а второе дерево -- на контрольной подргуппе. Разница предсказаний этих деревьев и даёт эффект от воздействия. Таким образом базовый элемент этого ансамбля можно отнести к методам непрямого оценивания. Сам ансамбль формируется методами Байесовского бутстрэппинга. В частности, считается что каждое наблюдение имеет вес $\theta_i \sim \mathtt{Exp}(1)$, где $\mathtt{Exp(1)}$ -- экспоненциальное распределение с параметром 1. Имея некоторую реализацию всех весов $\Theta$, рассматривается процесс, который может породить одно из имеющихся наблюдений $i$ с вероятностью $\theta_i$. Затем обучаются деревья, которые минимизирует ошибку предсказания на этом процессе при фиксированном $\Theta$. В силу того, что параметры этого процесса являются случайными величинами, то каждой реализации этих параметров соответствует свои деревья решений на тестовой и контрольной подвыборках, которые и формируют ансамбль. Данный подход предполагает очень большую выборку данных, так как наблюдения, за пределами имеющихся, не могут быть получены. Данный подход был успешно протестирован для анализа модификации сайта, предполагающего незначительное изменение клиентского опыта, и как следствие, небольшой средний эффект от такого воздействия. Авторы показывают, что такой подход способен успешно выявлять очень небольшую неоднородность эффекта от воздействия, в частности, авторы показывают, что даже на изменение размера картинки товара на сайте интернет магазина разные люди реагиуруют по разному.

\section{Сравнение качества работы некоторых методов}
\addtocru{section}{Сравнение качества работы некоторых методов}
\addtocen{section}{Experimental Evaluation of Some Models} 

Продемонстрируем, как различные методы выявления эффекта от воздействия ведут себя на синтетических данных. Будут рассмотрены 4 метода, по одному из каждой группы. Этот раздел не следует рассматривать как полноценное сравнение методов, которое выходит за рамки данной работы -- даже разработка методологии хорошего сравнения является ложной задачей. На разных данных будут лидировать разные методы. Соответственно, необходимо рассмотреть очень большой набор существенно различающихся данных как реальных, так и синтетических. Дальнейшая систематизация результатов также является сложной проблемой. Некоторые примеры сравнения различных методов могут быть найдены в~\cite{Guelman2015,SotysMichaandJaroszewicz2015,Devriendt2018}.

Для экспериментальной апробации некоторых методов была порождена случайным образом следующая выборка данных. В выборке присутствует три независимых переменных, две из которых являются численными, и распределены в соответствии с нормальным распределением $X_1,X_2 \sim N(0,1)$, а последняя является бинарной переменной, с вероятностью 60\% принимающая значение 1, а с вероятностью 40\% -- значение 0. 
В этой выбооке переменная отклика $Y$ будет бинарной переменной, принимающей положительные значения, когда $Y^* > 0$, где $Y^*$ рассчитываться по следующей формуле:
\begin{equation}
  Y^* = X_1 + X_3 \cdot T + \epsilon,\label{eq:exp-data}
\end{equation}
где $\epsilon \sim N(0,1)$ моделирует случайный шок. Здесь переменная $T$ задаёт связь между воздействием, если оно было ($T=1$) и $Y^*$. Таким образом, в этой выборке данных есть только одно воздействие ($T=1$), помимо базового ($T=0$). В реальной жизни, исследователь контролирует только значение этой переменной: кому-то на свой выбор он может назначить воздействие, а кому-то нет. В рамках данной работы рассматривается случайное назначение воздействие, которое зависит только от вероятности его назначения ($\PP{T=1}$). Для создания этой выборки данных будем назначать воздействие для отдельного наблюдения с вероятностью 50\%.
Переменная отклика $Y$ в этой выборке данных является бинарной равной 1, если $Y^* \geq 0$, и нулю в других случаях. 

Мы можем заметить, что в соответствии с формулой мы будем наблюдать положительный эффект от воздействия, когда $X_3 = 1$. И отсутствие эффекта от воздействия в противном случае. Действительно, эффект от воздействия для бинарной переменой $Y$, задаётся следующим образом: $\tau(X)=\PP{Y=1\mid T=1} - \PP{Y=1 \mid T=0}$. Тогда в соответствие с законом~(\ref{eq:exp-data}), $\tau(X_1,X_2,X_3)=\PP{X_1+X_3+\epsilon \geq 0} - \PP{X_1 + \epsilon \geq 0}$. Также отметим что в выборке данных есть переменная ($X_2$), не участвующая в этой формуле, что может осложнит задачу обучения.  В соответствие с законами распределения $X_1$, $X_2$, $X_3$, а также зная вероятность назначения воздействия ($\PP{T=1}$), можно создать выборку данных любого размера. Затем в соответствие с законом~(\ref{eq:exp-data}), можно рассчитать значение переменой отклика. Включим в создаваемую выборку данных 6000 наблюдений, каждое из которых связано с наличием или отсутствием воздействия.

В таблице~\ref{tbl:exp-data} представлено несколько наблюдений полученных в соответствие с законом~(\ref{eq:exp-data}). В этой таблице в первом столбце находятся уникальный идентификатор каждого наблюдения, который не используется для обучения модели, затем в следующих 5 столбцах идёт описание самой выборки данных, при этом во 2 и 3 столбцах находится бинарная переменная отклика $Y_g$ и переменная кодирующая какое воздействие ($t_g=1$ или $t_g=0$) было оказано на каждой наблюдение, и далее в 4--6 столбцах идут независимые переменные.  Помимо самой выборки данных в таблице~\ref{tbl:exp-data} также для каждого наблюдения показаны значения переменной $Y^*$, на основании которой вычисляется бинарная переменная $Y$, и случайный шок $\epsilon$ из закона~(\ref{eq:exp-data}), однако эти колонки не используется при обучении моделей оценки эффекта от воздействия, они представлены в таблице только для наглядности. 

\begin{table}
  \caption{Несколько наблюдений из синтетической выборки данных, на которой производилось сравнение различных моделей оценки индивидуальных эффектов от воздействия.}
  \label{tbl:exp-data}
  \centering
  \begin{tabular}{l|c|c||ccc||cc}
    \hline
    \hline
    &
    $Y_g$ & $t_g$ & $x_1^{(g)}$ & $x_2^{(g)}$ & $x_3^{(g)}$ & $Y_g^*$ & $\epsilon_g$ \\
    \hline
    $g_1$ &
    1 & 1 & -1.1 & -0.42 & 0 & 0.37 & 1.47 \\
    $g_2$ &
    1 & 0 & 0.71 & 0.37 & 1 & 1.97 & 1.26 \\
    $g_3$ &
    1 & 0 & 0.89 & -0.52 & 0 & 1.4 & 0.51 \\
    $g_4$ &
    1 & 1 & 1.45 & 0.04 & 1 & 2.89 & 0.44 \\
    $g_5$ &
    0 & 0 & -1.15 & -1.33 & 1 & -0.73 & 0.42 \\
    $g_6$ &
    1 & 0 & 1.89 & -0.46 & 0 & 2.7 & 0.81 \\
    $g_7$ &
    0 & 1 & 0.65 & 0.87 & 0 & -0.95 & -1.6 \\
    $g_8$ &
    1 & 1 & 1.38 & -1.59 & 1 & 2.08 & -0.3 \\
    $g_9$ &
    0 & 1 & -1.4 & -1.11 & 1 & -0.58 & -0.18 \\
    $g_{10}$ &
    1 & 1 & -0.26 & -0.39 & 1 & 1.83 & 1.09 \\
    \hline
    \hline
  \end{tabular}
\end{table}

Первый модель, качество которой проверяется на этих данных является подходом двух моделей, в котором в качестве базовой модели используется линейная модель линейной регрессии. Такие модели регрессии независимо оценивается на тестовой и контрольной выборке. 
В качестве второй модели рассмотрим модель Тиана~\cite{Tian2014}, который преобразовывает независимые переменные. 
Дальше рассмотрим один из методов построения модели деревьев решений, представленный в~\cite{Athey2016}. Последним методом будет метод \texttt{Uplift Random Forest}~\cite{Guelman2015} для построения случайного леса для оценки эффекта от воздействия. Эти методы являются достаточно популярными и представляют основные разновидности существующих методов. 

Первый и второй метод имеют простую реализацию и был реализован автором главы самостоятельно в пакете \texttt{SuDiTE}\footnote{\url{https://github.com/AlekseyBuzmakov/SuDiTE}} на языке R. Для третьего метода есть пакет для языка R\footnote{\url{https://github.com/susanathey/causalTree}} реализованный авторами работы~\cite{Athey2016}. В то же время последний метод реализован его авторами и доступен в пакете \texttt{uplift} также для R. Также в пакете \texttt{SuDiTE} есть средства для проверки методов выявления \SDuTE посредством перекрёстного тестирование\footnote{\textit{англ.} cross-validation}, которые будут описаны ниже. 

Все 4 метода оценивают эффект от воздействия. Для выделения конкретных подгрупп \SDuTE будем сравнивать оценку каждого метода эффекта от воздействия с нулём. И если метод предсказывает положительный эффект, то будем считать, что данный метод включается в подгруппу с положительным эффектом. Также добавим два синтетических метода, один из которых, назовём \texttt{ALL}. Метод \texttt{ALL} является наивным методом, которые всегда включает все наблюдения в \SDuTE, т.е. этот метод предполагает, что на всех наблюдениях есть положительный эффект от воздействия.
Второй синтетический метод назовём \texttt{BEST}. Это метод-оракул, который в качестве \SDuTE всегда выделяет правильные \SDuTE. На этих данных мы можем рассмотреть такой метод, потому что мы знаем закон~(\ref{eq:exp-data}), в соответствии с которым данные были порождены. Однако в реальной ситуации, этот метод принципиально не возможен и приводится в этом сравнении только как верхнюю границу качества, которая может быть получена на этих данных. В нашем случае в соответствие с законом~(\ref{eq:exp-data}) этот метод будет определять \SDuTE как наблюдения, для которых верно $x_3^{(g)}=1$.

Также нам необходимо повторить условия реального тестирования. В этих условиях есть одна выборка данных, и новые данные не могут быть порождены. Тогда на одной части выборки каждая модель должна быть обучена, а на другой проверена. Для этого мы разбиваем выборку данных на два множества: обучающее и проверочное. 
Обучающее множество используется для построения каждой модели, и на проверочном множестве обученные модели делают предсказание эффекта от воздействия и выделяют \SDuTE. Для выделенной подгруппы \SDuTE считается средний эффект от воздействия как $\EE{Y\mid T=1,\text{\SDuTE}} - \EE{Y\mid T=0, \text{\SDuTE}}$. Этот средний эффект от воздействия рассматривается как метрика качества работы соответствующей модели. Чем больше средний эффект от воздействия на \SDuTE, тем больше наблюдений с положительным эффектом туда попало, значит тем лучше модель.

Чтобы получить более надёжную оценку качества работы каждого метода, исходные данные 100 раз делятся на обучающую и проверочную выборки случайным образом. И, таким образом, для каждой модели будут известны 100 значений метрик качества их работы. Сведём все шаги сравнения различных моделей:
\begin{enumerate}
\item Порождение случайной выборки данных требуемого размера в соответствии с законом~(\ref{eq:exp-data});
\item Разбиение имеющейся выборки на обучающую и проверочную выборку данных случайным образом;
\item Выбор модели $M$ для её тестирования;
\item Обучение модели $M$ на обучающей выборке данных;
\item Предсказания эффекта от воздействия моделью $M$ для каждого наблюдения проверочной выборки данных;
\item Отбор наблюдений из проверочной выборки данных, для которых предсказаний эффекта от воздействия моделью $M$ больше нуля;
\item Для всего множества отобранных наблюдений проверочной выборки вычисляется средний эффект от воздействия в соответствии с $\EE{Y\mid T=1} - \EE{Y\mid T=0}$, этот средний эффект регистрируется как метрика качества модели $M$;
\item Шаги C--G выполняются для каждой тестируемой модели;
\item Шаги B--H повторяются заданное количества для возможности анализа распределения метрики качества, получаемой на шаге 7, для каждой модели.
\end{enumerate}

\begin{figure}[t]
  \centering
  \includegraphics[width=0.8\columnwidth]{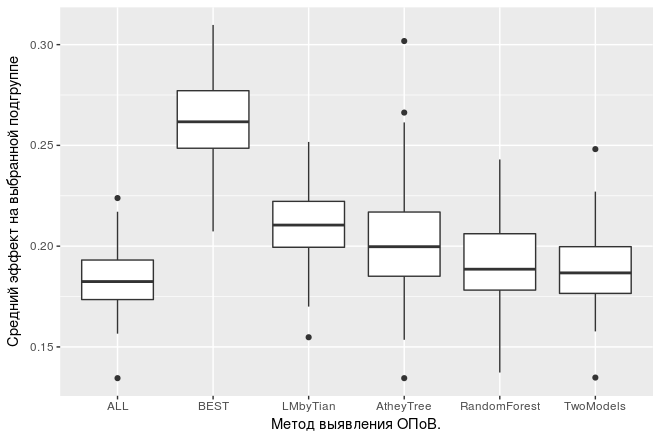}
  \caption{Сравнение некоторых методов выделения однородных подгрупп с точки зрения эффекта от воздействия на синтетических данных.\label{fig:comparison}}
\end{figure}

На рис.~\ref{fig:comparison} показаны распределения этих метрик качества для каждой из рассматриваемых моделей. Каждое распределение показывается посредством, так называемого <<ящика с усами>>, у которого границы прямоугольника (ящика) соответствуют квантилям уровня 25\% и 75\%, а линия по середине соответствует медиане.

На рис.~\ref{fig:comparison} мы можем заметить, что, как и ожидалось, метод \texttt{BEST}, который знает правильный ответ, выдаёт наибольшее значение по сравнению со всеми другими рассматриваемыми моделями соответствующей ему метрики качества.
Лучшим среди реальных моделей на этих данных является модель Тиана, которая при сравнимой вариативности с методом $\texttt{BEST}$ находится примерно по середине по качеству между \texttt{BEST} и $\texttt{ALL}$ (соответствует среднему эффекту от воздействия на проверочной выборке выборке данных).
Следом за моделью Тиана идёт метод построения дерева решений, вариативность предсказаний которого существенно выше, чем вариативность предсказания по модели Тиана. Далее, с ещё большей вариативностью, следует модель случайного леса. 
Одной из худших моделей на этих данных является подход двух моделей, качество работы которого, однако, не отличимо в статистическом смысле от качества работы случайного леса и при этом имеет меньшую вариативность.

Здесь нужно отметить, что такой порядок моделей в существенной степени объясняется тем, что модель генерации данных была линейной. Соответственно, линейные модели лучше всего подходят для моделирования таких данных. В реальных же ситуациях зависимость между независимыми переменными, переменной отклика и воздействием как правило существенно нелинейная, что в каждой конкретной ситуации приводит к своему упорядочиванию моделей, с точки зрения их качества работы на соответствующей выборке данных.

Также мы видим, что все 4 рассмотренных модели работают существенно хуже, чем метод-оракул \texttt{BEST}. 
При этом рассматриваемые синтетические данные являются относительно простыми, так как эффект от воздействия входит как линейный член в переменную отклика, а средний эффект составляет 27\% относительно бинарной переменной отклика $Y$. Это показывает, что методы выявления \SDuTE и методы оценки индивдуальных эффектов от воздействия, по-видимому, находятся на относительно ранней стадии своего развития, и многое ещё следует сделать, прежде чем будyт достигнуто качество предсказания, соответствующие качеству предсказания классических задачах машинного обучения.


\section{Заключение}
\addtocru{section}{Заключение} 
\addtocen{section}{Conclusion} 

В этой главе были рассмотрены существующие подходы к индивидуальной оценке эффекта от воздействия, и также методы выделения подгрупп однородных с точки зрения эффекта от воздействия. Данные методы применяются в различных областях от маркетинга до медицины. Их развитие может оказать большое влияние на многие области нашей жизни. 
Были рассмотрены 4 группы методов: методы не прямого оценивания эффекта от воздействия, методы сведения задачи оценки эффекта от воздействия к задачи регрессии, методы, основанные на деревьях решений, и ансамблевые методы. Некоторые из этих методов были экспериментально апробированы на синтетической выборке данных, на которой был задан достаточно большой эффект от воздействия. Однако даже этот эффект рассмотренные методы не смогли полностью выделить. Это означает, что необходимо создание новых эффективных методов оценки эффекта от воздействия для рассматриваемой области знаний.

Сложность разработки таких методов связана с несколькими аспектами. Во-первых, сама постановка задачи предполагает, что будет оцениваться средний эффект на некоторой подвыборке данных. При агрессивном выделение таких подвыборок, существенно повышается вероятность переобучения. 

Во-вторых, даже оценка качества таких методов вызывает затруднения, так как на индивидуальном уровне истинный эффект от воздействия неизвестен ни для одного наблюдения. Поэтому для оценки качества работы методов приходится использовать усреднение предсказаний модели по некоторой выборки данных достаточно большого размера.

В-третьих, практические задачи, в которых применяются методы оценки эффекта от воздействия, часто требуют оценки доверительных интервалов и статистической значимости для результатов и коэффициентов моделей, что затруднительно оценить для достаточно сложных моделей. 

\section*{Библиографический список}
\FloatBarrier 
\printbibliography
\end{refsection}  
\newpage 


\end{document}